\documentclass[12pt]{article}
\usepackage{geometry}	
\usepackage{graphicx} 
\usepackage{setspace}
\usepackage[justification=centering,font=small]{caption}
\usepackage{subcaption}
\usepackage{amsmath,amssymb}
\usepackage{appendix} 
\usepackage{placeins}
\usepackage[bookmarks=true]{hyperref} 
\usepackage{bookmark}
\usepackage{url}
\usepackage[small,compact]{titlesec}  
\usepackage{xcolor}	
\usepackage[authoryear,round]{natbib} 
\usepackage{amsthm}
\usepackage{threeparttable}
\usepackage{booktabs}
\usepackage{anyfontsize}
\usepackage{changepage}
\usepackage{float}

\allowdisplaybreaks[1]
\hypersetup{
	colorlinks,
	linkcolor={red!50!black},
	citecolor={blue!50!black}}

\theoremstyle{plain}
\newtheorem{thm}{Theorem}
\theoremstyle{definition}

\newtheorem{proposition}[thm]{Proposition}

\newtheoremstyle{indenteddefinition}
	{}
	{}
	{\hangindent=2em}
	{}
	{\bfseries}
	{.}
	{.5em}
	{}
\theoremstyle{indenteddefinition}

\newcounter{assumptiongroup}\stepcounter{assumptiongroup}

\newcommand{\E}{{ \mathbb{E} }}
\newcommand{\Prob}{{ \mathbb{P} }}
\newcommand{\argmax}[1]{{ \underset{ #1 }{ \mathrm{arg}\, \mathrm{max} } \ }}

\newcommand{\thetab}{{ \boldsymbol{\theta} }}
\newcommand{\Thetab}{{ \boldsymbol{\Theta} }}

\newcommand{\Xb}{{ \boldsymbol{X} }}
\newcommand{\xb}{{ \boldsymbol{x} }}
\newcommand{\Ib}{{ \boldsymbol{I} }}
\newcommand{\ib}{{ \boldsymbol{i} }}
\newcommand{\qb}{{ \boldsymbol{q} }}
\newcommand{\essb}{{ \boldsymbol{s} }}

\title{\vspace{-20mm} \Large Counting Defiers: A Design-Based Model of an Experiment Can Reveal Evidence Beyond the Average Effect\thanks{\scriptsize This paper is the culmination of a series of drafts \citep*{kowalski2019a,kowalski2019b,christy2024,christy2024a}. We extend special thanks to Jann Spiess for extensive regular feedback and to Charles Manski, Aleksey Tetenov, Toru Kitagawa, and Donald Rubin for encouraging us to use statistical decision theory and teaching us about it. We thank Guido Imbens for foundational feedback. Tory Do, Simon Essig Aberg, Jack Cavanaugh, Bailey Flanigan, Pauline Mourot, Srajal Nayak, Darion Phan, Sukanya Sravasti, Griffin Shufeldt, Matthew Tauzer, and Shuheng Zhang provided excellent research assistance.  For helpful comments, we thank Elizabeth Ananat, Don Andrews, Isaiah Andrews, Josh Angrist, Susan Athey, Victoria Baranov, Steve Berry, St\'ephane Bonhomme, Michael Boskin, Zach Brown, Kate Bundorf, Matias Cattaneo, Xiaohong Chen, Victor Chernozhukov, Janet Currie, Peng Ding, Guilherme Duarte, Pascaline Dupas, Brad Efron, Natalia Emanuel, Ivan Fernandez-Val, Chris Ferrie, Ashvin Gandhi, Michael Gechter, Andrew Gelman, Matthew Gentzkow, Andrew Goodman-Bacon, Sander Greenland, Florian Gunsilius, Andreas Hagemann, Sukjin Han, Jerry Hausman, Han Hong, Lawrence Katz, Daniel Kessler, Benedikt Koch, Michal Koles{\'a}r, Jonathan Kolstad, Ang Li, John List, Bentley MacLeod, Aprajit Mahajan, Jos{\'e} Luis Montiel Olea, Sendhil Mullainathan, Derek Neal, Andriy Norets, Matthew Notowidigdo, Amanda Pallais, Elena Pastorino, John Pepper, Demian Pouzo, Tanya Rosenblat, Andres Santos, Azeem Shaikh, Elie Tamer, Alexander Torgovitsky, Davide Viviano, Edward Vytlacil, Stefan Wager, Chris Walker, Christopher Walters, Thomas Wiemann, David Wilson, Zeyang Yu, seminar participants at Columbia, Harvard, MIT, the Norwegian School of Economics, Notre Dame, NYU, Princeton, Stanford, UCLA, UVA, the University of Chicago, the University of Michigan, the University of Pennsylvania, the University of Zurich, Yale, and conference participants at the Advances with Field Experiments Conference at the University of Chicago, the AEA meetings, the Bravo Center/SNSF Workshop on Using Data to Make Decisions, the Essen Health Conference, and the Y-RISE Evidence Aggregation and External Validity Conference. We thank Charles Antonelli, Bennett Fauber, Corey Powell, and Advanced Research Computing at the University of Michigan, as well as Misha Guy, Andrew Sherman, and the Yale High Performance Computing Center. This research was supported in part by the National Institute on Aging of the National Institutes of Health under Award Number R01AG089084. The content is solely the responsibility of the authors and does not necessarily represent the official views of the National Institutes of Health.   }}
\author{Neil Christy\thanks{\scriptsize University of Michigan}\; and Amanda Ellen Kowalski\thanks{\scriptsize University of Michigan and National Bureau of Economic Research}}
\date{June 21, 2026}

\begin{document}
	\maketitle
	\singlespacing
\begin{adjustwidth}{0.1in}{0.1in}
\begin{abstract}\fontsize{9pt}{12pt}\selectfont
{
\vspace{0.1in}
\noindent 

Using only a binary intervention and outcome and the design of the randomization within an experiment, we construct a design-based likelihood of the joint distribution of potential outcomes in the sample---the numbers of always takers, compliers, defiers, and never takers. We develop a visualization to show that samples with defiers can sometimes generate the data in more ways than samples without, yielding a higher likelihood. This likelihood can vary within the Fr\'{e}chet bounds, even though the traditional likelihood does not. Evidence is weak, but it exists, as we illustrate with health applications and our \texttt{dbmle} package.

\vspace{7mm}
}
\end{abstract}
\end{adjustwidth}
\newpage

\section{Introduction} \label{sec:intro}
We begin with a stylized example that demonstrates how a design-based model of an experiment can reveal evidence beyond the average effect. In Section \ref{sec:model}, we formalize the model and present a likelihood function for the joint distribution of potential outcomes in the sample, which we use to construct maximum likelihood estimates. In Section \ref{sec:applications}, we use our design-based likelihood in two real-world health applications. Section \ref{sec:contributions} discusses our contributions to the literature. Section \ref{sec:implications} presents implications for econometric and applied research.

\subsection{Visualization of a Stylized Example:  A Design-Based Model Can Reveal Evidence Beyond the Average Effect } \label{sec:example}

You are a doctor at a rural health clinic, and you have six patients who are smokers. You would like them to quit smoking, so you consider offering them a payment contingent on quitting.  You are optimistic that the payments would induce some patients to quit.  However, your psychologist friend warns you that payments can crowd out intrinsic motivations \citep*{gneezy2000}. Your economist friend tells you that traditional economic models assume away the possibility \citep*{bjorklund1987, imbens1994, heckman1999, vytlacil2002}.  But you have taken an oath to ``do no harm.”  Even if the payment works on average, that is little comfort if some of your patients would actually be harmed. The more patients who would quit otherwise but would not because of your payment, the worse you would feel. For patients who would quit regardless, your concern is more modest---you would rather not waste the payments. Because the payment is contingent on quitting, you would not waste payments on any patients who would remain smoking regardless of the payment, but as their doctor, you would still feel bad that your intervention cannot reach them.   You would really like to know the count of patients of each of the four types. 

You know that experiments are a tool to reveal causal evidence on the average effect.  You also know that with traditional inference, you can obtain Fr\'{e}chet bounds \citep*{boole1854, hoeffding1940, frechet1957} on the distribution of effects in the population from which the smokers in your clinic were drawn.  However, since your responsibilities are to your own patients, you do not focus on the population. You would like to estimate the full distribution of effects in your fixed sample of patients.

You plan an experiment, and you are deliberate about the \emph{design} of your randomization. You choose a ``completely randomized’’ design in which exactly three patients will be assigned to the payment intervention.  You implement this design by using a computer to effectively pick three names out of a hat, without replacement.  

You run the experiment, and the results are promising in terms of the average effect. Two of the three patients in intervention quit smoking; only one of the three patients in control quits.  The intervention mean is 2/3, and the control mean is 1/3, so the average effect of the payment intervention is an increase in the quit rate of 1/3. 

\textit{Why} did you observe the data you observed?  What would you have seen under the other possible randomized assignments? Can the experiment reveal any evidence about effect heterogeneity? In more technical terms, what is the joint distribution of ``potential outcomes” \citep*{neyman1923,rubin1974, rubin1977} in the sample? Counterfactual questions like these have attracted recent interest in the study of causal inference \citep*{gelman2013, pearl2018, imbens2020, dawid2022}.  

We assume, as in design-based inference, that the potential outcomes of each person are fixed, and therefore the joint distribution of potential outcomes in the sample is fixed, before randomization.  Following \citet*{angrist1996}, we classify patients into types based on their potential outcomes in intervention and control:  “always takers” take up the desired health behavior---they quit smoking---in intervention and control (effect=0); “compliers” take up in intervention but not control (effect=1); “defiers” do not take up in intervention but take up in control (effect=-1); and “never takers” do not take up in intervention or control (effect=0). We specify the joint distribution of potential outcomes with the numbers of always takers, compliers, defiers, and never takers in the sample.  The joint distribution of potential outcomes provides more information than the average effect; it provides the full distribution of effects and disambiguates the zero effects of always takers from those of never takers.  

\begin{figure}[htp!]
	\centering
	\caption{In an Experiment with Six Patients, the Likelihood Varies Among Joint Distributions of Potential Outcomes with the Same Marginal Distributions as the Data (2/3 Takeup in Intervention, 1/3 Takeup in Control) and is Maximized with Four Compliers and Two Defiers}
	\makebox[\textwidth][c]{
        \hspace{0.00\textwidth}
        \includegraphics[width=0.93\textwidth]{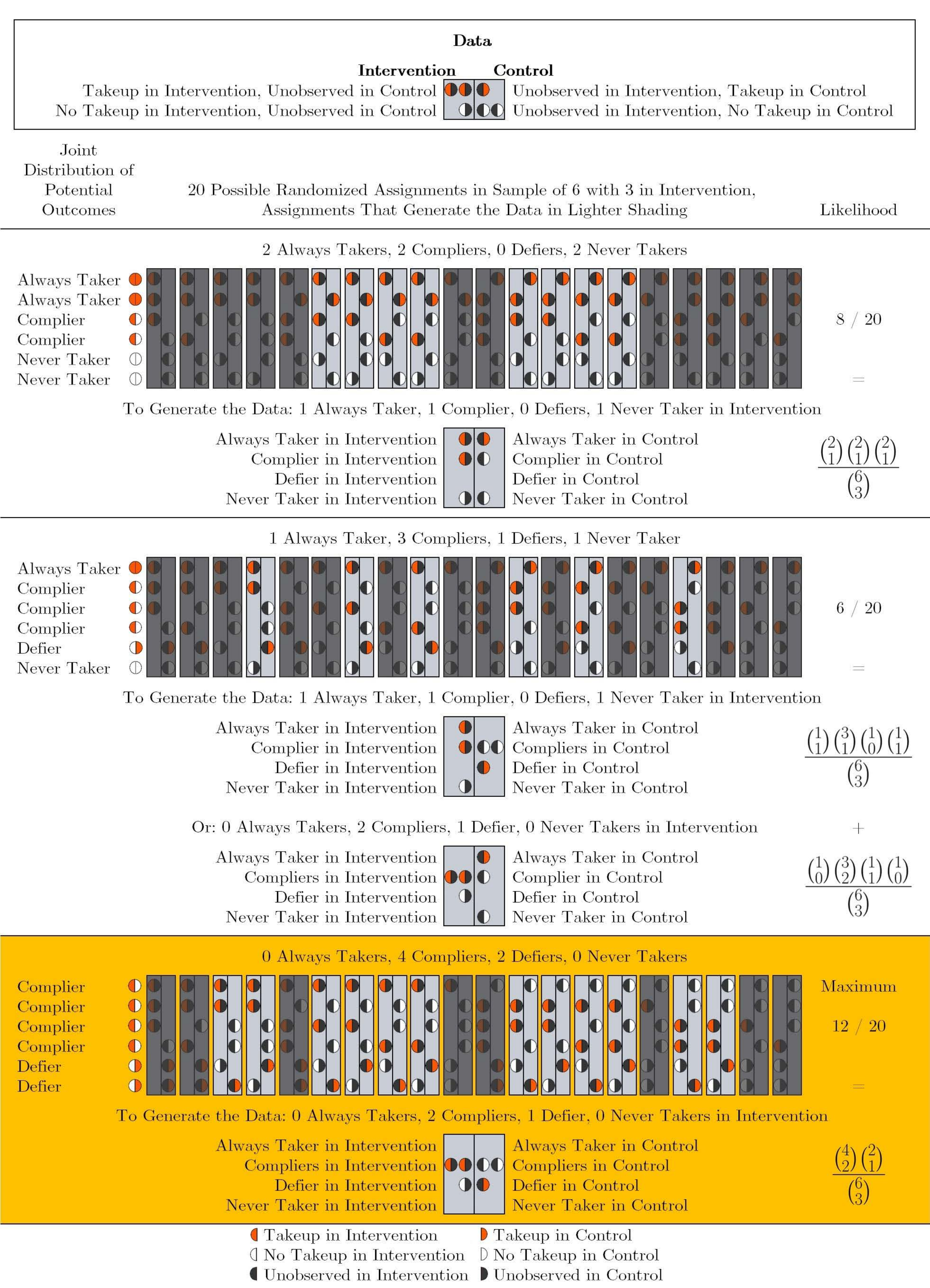}
        }
	 \label{fig:six}
\end{figure}

We demonstrate that a design-based model of your experiment can reveal evidence on the numbers of always takers, compliers, defiers, and never takers in your clinic, beyond the evidence revealed by the estimated average effect and the resulting Fr\'{e}chet bounds. We develop a novel visualization of potential outcomes to illustrate how in Figure \ref{fig:six}. We represent each of the six patients in the sample with a circle. The left half of each circle represents the potential outcome in the payment intervention, and the right half represents the potential outcome in the no payment control. Orange represents “takeup” of the desired health behavior---quitting smoking; white represents “no takeup,” and grey represents ``unobserved.''

Rectangles with lines down the middle represent experiments with intervention on the left and control on the right.  Inside each experiment, assignment to intervention reveals the potential outcome in intervention---the left of the rectangle reveals the left of the circle, which can be orange for takeup or white for no takeup---but the potential outcome in control is unobserved, so it is grey.  Similarly, assignment to control reveals the potential outcome in control---the right of the rectangle reveals the right of the circle, which can be orange for takeup or white for no takeup---but the potential outcome in intervention is unobserved, so it is grey. 

The data from your experiment, depicted in the box at the top of Figure \ref{fig:six}, show takeup among two of three patients in intervention and one of three patients in control.  Therefore, the estimated marginal distribution of the potential outcome in intervention is 2/3 takeup, and the estimated marginal distribution of the potential outcome in control is 1/3 takeup.  These two marginal distributions define the ``estimated Fr\'{e}chet set.’’ We refer to all joint distributions of potential outcomes with the same marginal distributions as members of the same “Fr\'{e}chet set.”  

The three rows of Figure \ref{fig:six} give the three joint distributions of potential outcomes with the same marginal distributions as the estimated Fr\'{e}chet set: half circles in which the potential outcomes are unobserved in the data are filled in so that they have the same marginal distributions as the data.  Four of the six circles have orange on the left (takeup occurs among 2/3 of patients in intervention), and two of the six circles have orange on the right (takeup occurs among 1/3 of patients in control).  As shown in the three rows, there are three ways to combine the colored half circles into colored full circles, each yielding a different number of defiers (circles with white on the left and orange on the right).  This depiction demonstrates how the estimated Fr\'{e}chet set determines the estimated Fr\'{e}chet bounds on defiers, which extend here from a lower bound of 0 to an upper bound of 2. Is there any variation in the design-based likelihood across the joint distributions of potential outcomes in the three rows?

The joint distribution of potential outcomes in the first row is consistent with the \citet*{imbens1994} assumption of “monotonicity” in the sample, whereby there can be compliers or defiers, but not both.  Since the estimated effect of 1/3 is positive, you assume away defiers. You then deduce that the person who takes up in control must be an always taker and the person who does not take up in intervention must be a never taker.  Assuming that the count of each type is the same in each arm, the sample includes two always takers and two never takers.  The remaining two patients must be compliers, so the joint distribution of potential outcomes consistent with the monotonicity assumption includes two always takers, two compliers, and two never takers.  However, you are uncomfortable assuming monotonicity because you are concerned that there could be defiers, and you would prefer to rely on evidence.    

The joint distributions of potential outcomes in the next two rows include defiers.  The one in the second row includes one always taker, three compliers, one defier, and one never taker.  The one in the third row includes four compliers and two defiers.  There is a rationale in the literature for assuming the distribution in the second row---it has one defier, the midpoint \citep*{li2019} of the Fr\'{e}chet bounds on defiers.  It can therefore seem surprising that evidence yields a rationale for assuming the distribution in the third row with two defiers.

Each row in Figure \ref{fig:six} depicts evidence from a ``design-based model’’ that combines the design of the randomization with a model of potential outcomes to specify a data generating process and use it to generate all possible data. As depicted in the top of each row, there are 20 possible randomized assignments of six patients that result in three in intervention ($6 \text{ choose } 3 = 20$).  By the design of your experiment, each of these possible randomized assignments has the same probability, so you can make the assumption that they each have the same probability very compelling through careful implementation. 

Within each main row, each of the six patients in the experiment has a separate row.  The person’s type and randomized assignment to intervention or control determines the data the experiment reveals, which can be counterfactual to what you observed.  The assignments that generate the data you observed---2/3 takeup in intervention and 1/3 takeup in control---are in lighter shading. 

The number of randomized assignments that generate the data you observed gives the numerator of the design-based likelihood of the joint distribution of potential outcomes, reported in the third column.  Paraphrasing the baby board book ``Statistical Physics for Babies'' \citep*{ferrie2017}, which provides style inspiration for our visualization, physicists refer to the number of ways that you could have seen what you have seen as ``entropy.”  The design-based likelihood is equal to the entropy---the number of assignments that generate the data (shown in lighter shading)---divided by the number of possible assignments.  The first row, with zero defiers, generates the data in eight of 20 possible assignments, so the likelihood is $8/20=40\%$.  The second row, with one defier, generates the data in six of 20 possible assignments, so the likelihood is $6/20=30\%$.  The last row, with two defiers, generates the data in 12 of 20 possible assignments, so the likelihood is $12/20=60\%$.

Why do some joint distributions of potential outcomes generate the data in more randomized assignments than others and therefore have higher likelihoods than others?  The bottom half of each row sorts the six patients by type and separately shows how the always takers, compliers, defiers, and never takers must be randomized into intervention and control to generate the data.  In the first row, there are three  types---always takers, compliers, and never takers---and there must be a balance between intervention and control within each type to generate the data.  In the second row, there are four  types---always takers, compliers, defiers, and never takers---and there must be an imbalance between intervention and control within each type to generate the data.  In the third row, there are only two types---compliers and defiers---and there must be a balance between intervention and control within each type to generate the data.

Likelihoods are higher when the patients who generate the data are as similar as possible in two senses: they belong to fewer types, and they have a counterpart of the same type in the opposite arm such that those of the same type are balanced between intervention and control.  The patients who generate the data are most similar in the third row: they belong to the fewest number of types (only two), and those of the same type are balanced between intervention and control.  Comparison of the first and third rows illustrates how having fewer types can increase the likelihood. In the first row, there are two ways to randomize one of two always takers into intervention ($2 \text{ choose } 1=2$), two ways to randomize one of two compliers into intervention ($2 \text{ choose } 1=2$), and two ways to randomize one of two never takers into intervention ($2 \text{ choose } 1=2$), so there are eight assignments that generate the data ($2 \times 2 \times 2=8$).  In the third row, there are six ways to randomize two of four compliers into intervention ($4 \text{ choose } 2=6$) and two ways to randomize one of two defiers into intervention ($2 \text{ choose } 1=2$), so there are 12 assignments that generate the data ($6 \times 2=12$).  Joint distributions of potential outcomes other than those in Figure \ref{fig:six} can generate the data; the true joint distribution of potential outcomes need not preserve the estimated marginal distributions.  For example, the likelihood that the true average effect is zero and your sample contains only always and never takers is $9/20=45\%$.  However, by exhaustive grid search, you determine that the global maximizer of the likelihood is the one depicted in the third row, which contains four compliers and two defiers. 

As a doctor considering the distribution with the maximum likelihood, you are thrilled that it implies that the payment intervention would induce four of your six patients to quit smoking, which is even more promising than what you would conclude based on the average effect and a monotonicity assumption.  You are also encouraged that it implies that you do not waste any payments on patients who would have quit regardless and that there are no patients who would continue smoking regardless.  However, you are concerned that it implies that the payment would induce two patients who would otherwise have quit to continue to smoke.

\section{A Design-Based Model of an Experiment} \label{sec:model}

\subsection{A Design-Based Model Implies a Likelihood for the Joint Distribution of Potential Outcomes in the Sample}

Here, we formalize the design-based model of an experiment presented in our stylized example.  A sample consists of $n$ subjects, each of whom has a binary potential outcome in intervention $Y_I \in \{0, 1\}$ and a binary potential outcome in control $Y_C \in \{0, 1\}$, where $1$ represents takeup and $0$ represents no takeup.  A subject's observed outcome depends only on their own potential outcomes and their inclusion in the intervention or control arm, ruling out network-type effects through a ``no interference’’ \citep*{cox1958} or ``stable unit treatment value’’ \citep*{rubin1980} assumption.
  
Subjects receive a random assignment $Z$ to intervention ($Z=I$) or control ($Z=C$), which is independent of a subject’s potential outcomes $(Y_I, Y_C)$.  The observed outcome $Y$ is:
\begin{align}
	Y = \mathbf{1}_{\{Z=I\}}(Y_I) + \mathbf{1}_{\{Z=C\}}(Y_C),
\end{align}
where $\mathbf{1}_{\{\cdot\}}$ is the indicator function.

Subjects in the sample belong to one of four types or ``principal strata,’’ defined by combinations of potential outcomes \citep*{frangakis2002}.   Let $\theta_{11}$ represent the number of always takers $(Y_I = 1, Y_C = 1)$, $\theta_{10}$ the number of compliers $(Y_I = 1, Y_C = 0)$, $\theta_{01}$ the number of defiers $(Y_I = 0, Y_C = 1)$, and $\theta_{00}$ the number of never takers $(Y_I = 0, Y_C = 0)$.  The collection of these four counts $\thetab = (\theta_{11}, \theta_{10}, \theta_{01}, \theta_{00})$ constitutes the joint distribution of potential outcomes in the sample, which we write as counts rather than shares to ease the following likelihood notation.  The experimental data $\Xb = (X_{I1}, X_{I0}, X_{C1}, X_{C0})$ consist of the counts of subjects who take up in intervention $X_{I1}$, who do not take up in intervention $X_{I0}$, who take up in control $X_{C1}$, and who do not take up in control $X_{C0}$.   

We adopt a “design-based” approach, in which the joint distribution of potential outcomes $\thetab$ is fixed, but unknown, and all randomness in the experimental data $\Xb$ comes from the random assignment of subjects to intervention or control.  We focus on the performance of decision rules within the finite sample, rather than on their asymptotic properties. As shown in \ref{sec:A_likelihood_design}, randomization implies a distribution for the experimental data $\Xb$ given the joint distribution of potential outcomes $\thetab$, yielding a likelihood expression.  For an experiment with Bernoulli randomization (also known as simple randomization or independent and identically distributed randomization) or a completely randomized experiment, the likelihood is proportional to:
\begin{align}
	\mathcal{L}( \thetab \mid \xb )
	&\propto \sum_{j \in \mathcal{I}(\xb, \thetab)}
	\binom{ \theta_{11} }{ j } \nonumber\\*
	&\qquad \qquad \times
		\binom{ \theta_{10} }{ x_{I1} - j } \nonumber\\*
	&\qquad \qquad \times
		\binom{ \theta_{01} }
			{ \theta_{11} + \theta_{01} - x_{C1} - j } \nonumber\\*
	&\qquad \qquad \times
		\binom{ \theta_{00} }
			{ x_{I0} + x_{C1} + j - \theta_{11} - \theta_{01}}, \label{eq:likelihood}
\end{align}
where the set $\mathcal{I}(\xb, \thetab)$ restricts $j$ such that the binomial coefficients remain well defined.  This expression appears in \citet*{copas1973}.  The likelihoods we derive in the stylized example of Section \ref{sec:intro} are also proportional to (\ref{eq:likelihood}).  For the first distribution with zero defiers and the third distribution with two defiers in Figure \ref{fig:six}, there is a single term in the likelihood summation.  For the second distribution with one defier, there are two.

\subsection{The Design-Based Likelihood Can Vary with the Number of Defiers within the Fr\'{e}chet Bounds Determined by the Estimated Marginal Distributions of Potential Outcomes} \label{sec:frechet}

In the previous section, we present a likelihood for the joint distribution of potential outcomes in the sample.  Each joint distribution of potential outcomes implies marginal distributions of potential outcomes in the intervention and control arms.  The marginal distribution of the potential outcome in intervention is the number of subjects for whom $Y=1$ in intervention:
\begin{align*}
	\theta_{1 \bullet} \equiv \theta_{11} + \theta_{10}.
\end{align*}
The marginal distribution of the potential outcome in control is the number of subjects for whom $Y=1$ in control:
\begin{align*}
	\theta_{\bullet 1} \equiv \theta_{11} + \theta_{01}.
\end{align*}

Given a pair of marginal distributions of potential outcomes in intervention and control, there are numerous joint distributions consistent with them. \cite*{boole1854}, \cite*{hoeffding1940}, and \cite*{frechet1957} derive bounds on the possible copulas connecting marginal distributions of random variables into joint distributions.  By applying these bounds to a pair of marginal distributions of potential outcomes, we can determine the set of all joint distributions of potential outcomes consistent with the pair of marginal distributions.  We refer to the set of joint distributions of potential outcomes consistent with a given pair of marginal distributions as the ``Fr\'{e}chet set’’ $\mathcal{F}(\theta_{1 \bullet}, \theta_{\bullet 1};n)$ of those marginal distributions:
\begin{align*}
	\mathcal{F}(\theta_{1 \bullet}, \theta_{\bullet 1};n) 
	= \bigg\{ \thetab \in \mathbb{Z}^4_+ 
	:\,& \theta_{11}+\theta_{10} = \theta_{1\bullet},\ \theta_{11} + \theta_{01} =\theta_{\bullet 1},\ 
    \sum_{i,j \in \{0,1\}} \theta_{ij} = n \bigg\}.
\end{align*}
In Figure \ref{fig:six}, the three rows comprise the Fr\'{e}chet set for $\theta_{1 \bullet} = 4$ and $\theta_{\bullet 1} = 2$. We index the distributions in a Fr\'{e}chet set by their number of defiers, $\theta_{01}$, which can take values in the range
\begin{align}
	\max \{0, -(\theta_{1 \bullet} - \theta_{\bullet 1})\} \leq \theta_{01} \leq \min \{\theta_{\bullet 1}, n - \theta_{1 \bullet} \}. \label{eq:frechet_bounds}
\end{align}
These inequalities provide the so-called ``Fr\'{e}chet bounds’’ on the number of defiers.  

We can form estimates of the marginal distributions of potential outcomes in intervention and control using the observed takeup rates in intervention and control:
\begin{align}
	\hat{\theta}_{1 \bullet}(\xb) = n
	\left(\frac{x_{I1}}{x_{I1} + x_{I0}}\right),
    \qquad
	\hat{\theta}_{\bullet 1}(\xb) = n
	\left(\frac{x_{C1}}{x_{C1} + x_{C0}}\right).  \label{eq:marginal_est}
\end{align}
In our design-based model of an experiment, these are unbiased estimators of the sample marginal distributions of potential outcomes for a completely randomized experiment.  In a standard sampling-based model of an experiment, dividing by $n$ yields consistent estimates of the population marginal distributions of potential outcomes.

It is well-known that, in the sampling-based model of an experiment, the takeup counts in intervention and control contain no information about the joint distribution of potential outcomes in the population beyond its marginal distributions.  In \ref{sec:A_sampling}, we recreate this result by showing that the likelihood of the population-level joint distribution of potential outcomes is constant, conditional on the population-level marginal distributions.  In other words, the likelihood function is flat within every population-level Fr\'{e}chet set.

\begin{figure}[bthp]
	\centering
	\caption{The Likelihood Varies with the Number of Defiers within the Estimated Fr\'{e}chet Set, And The Smallest Collection of Likelihoods to Contain 95\% of the Mass Excludes 105 to 116 Defiers (in Lighter Shading) and the Midpoint}
	\includegraphics[width=\textwidth]{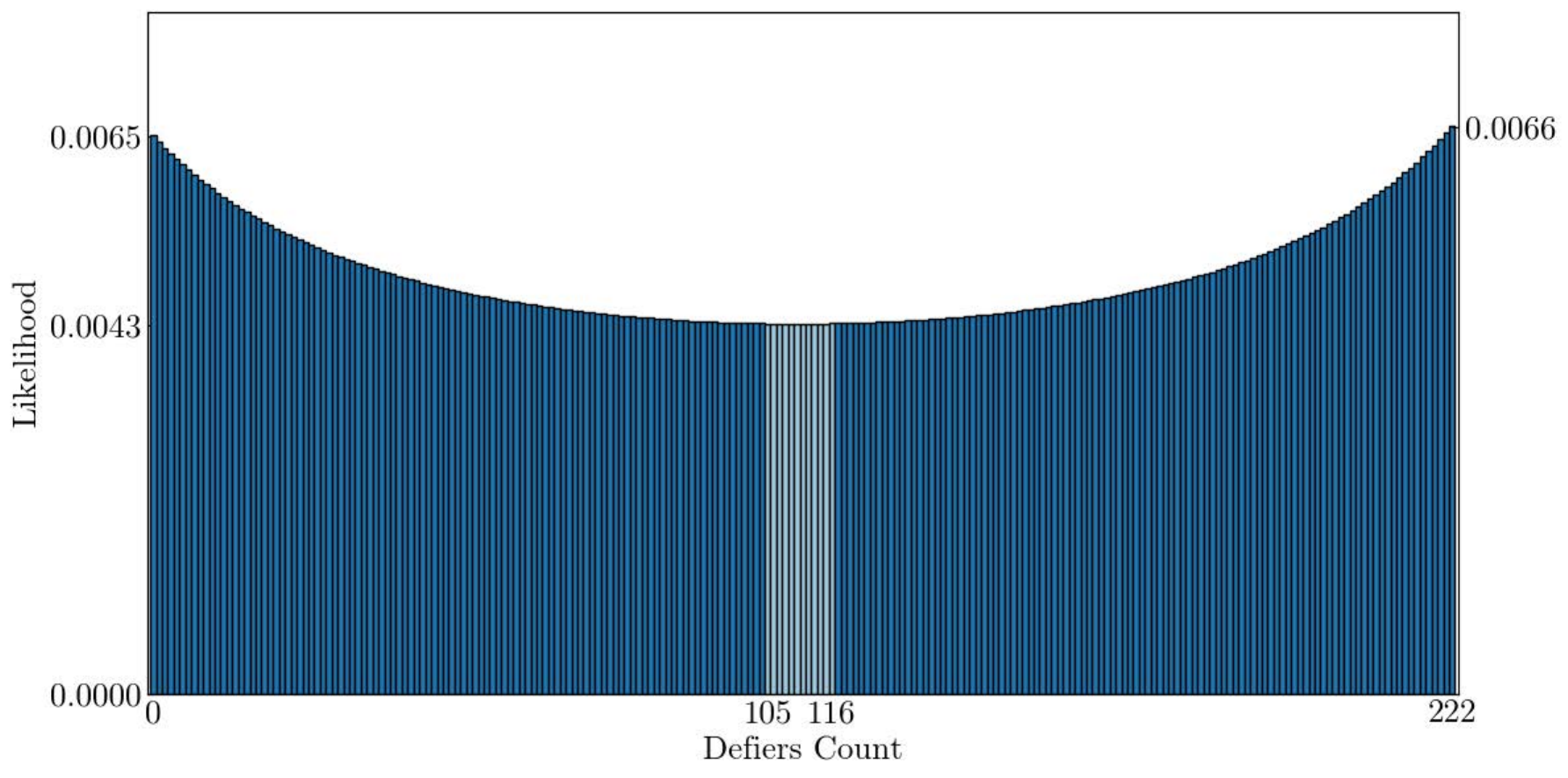}
	\label{fig:frechet}
\end{figure}

In our design-based model of an experiment, however, the data can contain information about the sample joint distribution of potential outcomes beyond the sample marginal distributions.  That is, the likelihood function can vary with the number of defiers within sample-level Fr\'{e}chet sets. Figure \ref{fig:frechet} plots the likelihood values of each joint distribution of potential outcomes in the estimated Fr\'{e}chet set formed by the data $(x_{I1}, x_{I0}, x_{C1}, x_{C0})= (154, 152, 111, 195)$, a hypothetical version of the first empirical example we discuss in Section \ref{sec:applications}.  We index the joint distributions in this Fr\'{e}chet set by their number of defiers along the horizontal axis.  The height of the bar is the likelihood of the joint distribution in the Fr\'{e}chet set with the specified number of defiers.  Within this Fr\'{e}chet set, the likelihood is highest at the distribution with 222 defiers, the upper Fr\'{e}chet bound on defiers.  As the number of defiers increases, the likelihood initially decreases, and then increases. If we normalize likelihoods in this Fr\'{e}chet set to sum to one, the smallest collection of likelihoods to contain 95\% of the mass excludes 105 to 116 defiers, indicated by the lighter shading. Notably, it excludes 111 defiers, the midpoint \citep*{li2019} of the Fr\'{e}chet set.

\subsection{The Design-Based Likelihood Enables Maximum Likelihood Estimation} \label{sec:rule}

One way to exploit the curvature in the design-based likelihood is to estimate the joint distribution of potential outcomes with maximum likelihood:
\begin{align*}
    \widehat{\thetab}_{MLE}(\xb) = {\arg \max}_{\thetab} \mathcal{L}(\thetab \mid \xb ).
\end{align*}
There are finitely many vectors of four integers that sum to the number of participants in the experiment $n$, so at least one value $\thetab$ maximizes the likelihood.  We solve this optimization problem through an exhaustive grid search over the finitely many joint distributions of potential outcomes consisting of $n$ subjects.

While maximum likelihood estimation is common in practice, we also provide three justifications in \ref{sec:justifications}, which begins in \ref{sec:A_bayes_prelim} by characterizing our estimates as the output of a statistical decision rule using the framework of statistical decision theory. First, we demonstrate that our estimates vary systematically with the data, as we show in Figure \ref{fig:largeheatmap} and discuss in \ref{sec:vary}.  Second, in \ref{sec:A_bayes}, we further justify our estimates as Bayes optimal under the appropriate conditions, including a uniform prior and a utility function that rewards ``being correct'' (a decision maker with a different utility could derive a different Bayes optimal rule from the design-based likelihood, which we demonstrate in Proposition \ref{prop:MSE} of  \ref{sec:A_bayes_mse} by deriving a Bayes optimal minimum mean squared error rule). Bayes optimality lets us quantify the gains from our rule: in a sample size of 200, the Bayes expected utility of our proposed maximum likelihood rule is 1.79 times higher than that of a Fr\'{e}chet rule and 1.56 times higher than that of a monotonicity rule, and the relative performance of our rule increases with the sample size.  A Bayesian perspective also allows us to quantify the strength of evidence by constructing credible sets.  Third, in \ref{sec:entropy}, we discuss how the maximum likelihood estimate is optimal based on the  principle of maximum entropy \citep*{jaynes1957a, jaynes1957b}, which ``assumes the least" \citep*{jaynes1968} about the joint distribution of potential outcomes conditional on the observed data.

\section{Applications: Counting Defiers in Health Care} \label{sec:applications}

We consider two published experiments with positive, statistically significant average effects on takeup of desired health behaviors and plausible defiers.  Figure \ref{fig:largeheatmap} summarizes the results from all possible empirical applications in experiments with given sizes, illustrating empirical conditions under which our maximum likelihood estimates yield defiers.  It is predictable from the data for each application and the patterns in Figure \ref{fig:largeheatmap} that our first application yields no defiers and our second application yields defiers, which we confirm with exhaustive grid search.  Our focus here is to provide examples to illustrate how applied researchers can present and interpret quantitative results and to provide substantive insights.  The top of Table \ref{tab:results} summarizes the context and design for both applications. The statistics in the table can be calculated using our Python- and Stata-compatible \texttt{dbmle} package, documented at \href{https://pypi.org/project/dbmle/}{https://pypi.org/project/dbmle/} \citep*{christy2025dbmle}.

\begin{table}[htbp]
	\caption{Standard Statistics and Proposed Design-Based Maximum Likelihood Estimates}
	\includegraphics[width=\linewidth]{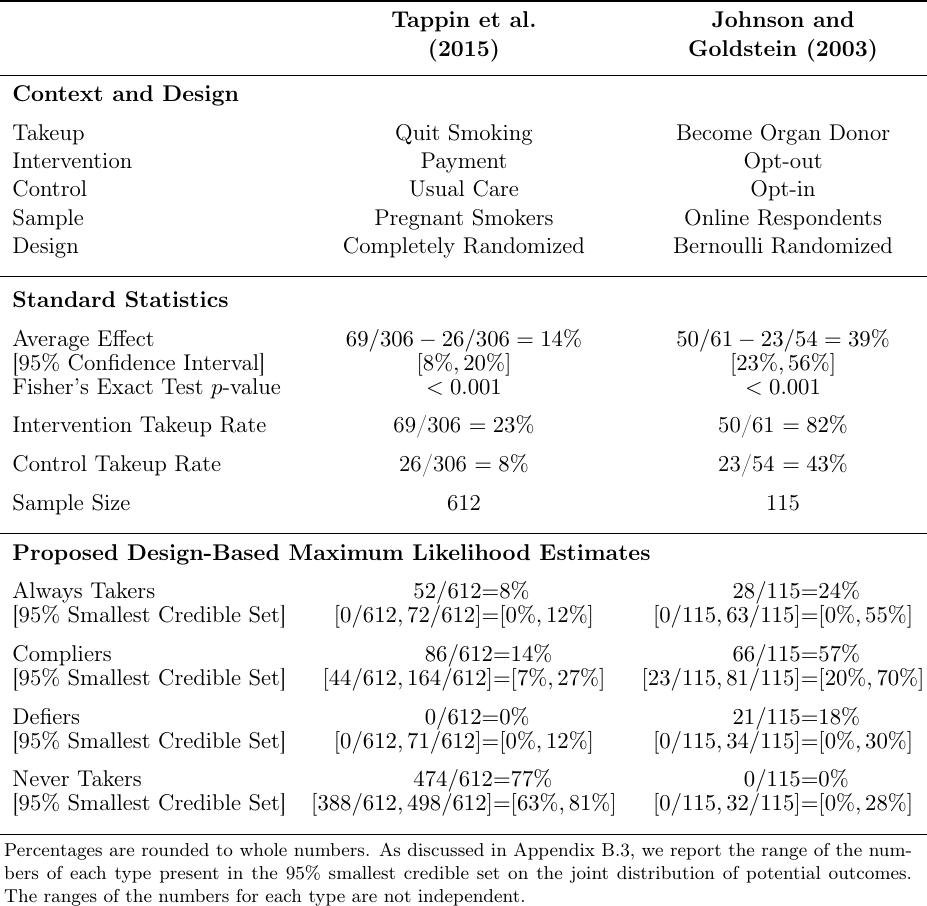}
	\label{tab:results}
\end{table}

As in our stylized example, in \cite{tappin2015}, the definition of ``takeup’’ is to quit smoking, the intervention involves payment, and the design is completely randomized with half assigned to intervention.  The sample includes 612 pregnant smokers. Monotonicity assumptions have a long history in the context of interventions that induce pregnant women to quit smoking.  In a study prominently discussed by \citet*{angrist1996}, \citet*{permutt1989} propose an early monotonicity assumption requiring that women who would quit smoking in control would also quit smoking in an arm that received a multifaceted intervention.  However, the intervention in \cite{tappin2015} involves payment, and as the psychologist friend warned in the introduction, literature since the monotonicity assumption was proposed expresses concern that payments can backfire in many contexts, including on-time pickup of children from day care \citep*{gneezy2000}, blood donation \citep*{mellstrom2008, lacetera2010}, and vaccination \citep*{schneider2023}.  Defiers who would quit smoking in control but not in intervention are plausible in \cite{tappin2015}  because a payment could backfire by crowding out the intrinsic motivations of pregnant women to quit smoking based on moral ideals and health goals.   

The average effect of the payment relative to usual care is a statistically significant 14 percentage point increase in the smoking quit rate on a base of 8\% in control.  Our maximum likelihood estimates reveal evidence beyond the average effect: $52/612=8\%$ always takers (effect=0), $86/612=14\%$ compliers (effect=1), $0/612=0\%$ defiers (effect=-1), and $474/612=77\%$ never takers (effect=0).  The weighted average effect across the four types ($8 \times 0+14 \times 1+0 \times (-1)+77 \times 0$) is equal to the estimated average effect of 14 percentage points, which is also equal to the share of compliers.  We would obtain the same results if we assumed monotonicity in the sample and deduced the remaining counts by assuming balance between intervention and control. Therefore, our estimates provide evidence in favor of a joint distribution of potential outcomes that satisfies monotonicity without assuming it.  

Returning to the concerns of the doctor in the introduction, the maximum likelihood estimates are reassuring.  No patients would be harmed by payment---the doctor's greatest concern. The 8\% who are always takers represent a modest waste, although the doctor is not too disappointed about giving payments to pregnant women who quit smoking. The 77\% of patients who are never takers cannot be reached by the intervention, which is disappointing but not costly given that payment is contingent on quitting.  The 14\% who are compliers are exactly the patients the doctor hoped to help.

Our design-based perspective affects the interpretation of our results.  The doctor in the introduction cares about the patients in the sample, and our design-based likelihood is informative because it conveys information about the sample at hand.  However, if the doctor were interested in the effects of a payment policy in the population from which the sample was drawn, the sampling-based likelihood we derive in \ref{sec:A_sampling} would reveal no information about defiers in the population beyond the Fr\'{e}chet bounds---at least, not without a departure from standard sampling assumptions, which we explore in \ref{sec:A_finite}.

We emphasize that even though our maximum likelihood estimate of the joint distribution of potential outcomes in the sample includes no defiers, considerable uncertainty remains. As discussed in \ref{sec:A_credible}, we compute a 95\% smallest credible set on the joint distribution of potential outcomes, and we report the ranges for the numbers of always takers, compliers, defiers, and never takers in that set in Table \ref{tab:results}.  The 95\% smallest credible set includes distributions with 0 to 71 defiers, representing 0\% to 71/612=12\% of the sample. The estimate that minimizes mean squared error, which we derive in Proposition \ref{prop:MSE} of \ref{sec:A_bayes}, lies in the interior of this range, at 4.4\% (Table \ref{tab:mse}). The magnitudes of these values highlight that evidence on defiers is weak when the estimated average effect is positive.

Building on the maximum likelihood estimate of the joint distribution of potential outcomes, researchers could go on to assume monotonicity, then proceed as usual assuming balance between intervention and control.  With additional data on covariates, they could calculate average characteristics of always takers, compliers, and never takers \citep*{imbens1997, katz2001, abadie2002, abadie2003}.  With additional data on a second stage outcome like birth weight, they could interpret the resulting instrumental variable estimate as a local average treatment effect on compliers \citep*{imbens1994}.  

Note that the joint distribution of potential outcomes can change dramatically even when the average effect does not.  If the control takeup rate in \cite{tappin2015} were $111/306=36\%$ instead of 8\%, with the same average effect of 14 percentage points (takeup among an additional 43 people in intervention), the maximum likelihood estimates would include 36\% defiers rather than zero. Note that these are the same data used for Figure \ref{fig:frechet} at the end of Section \ref{sec:frechet}. Our second application illustrates the estimation of a positive count of defiers with real data.  

Our second application, \citet*{johnson2003}, is a high-profile experiment in which a simple change from an opt-in default to an opt-out default increases organ donation almost twofold from 43\% in control to 82\% in intervention.  Results from \citet*{johnson2003} have inspired state policy changes to encourage life-saving organ donations to combat chronic organ shortages \citep*{kessler2025}.  However, in a recent experiment by \citet*{kessler2025}, in which takeup represents actually becoming an organ donor instead of hypothetically becoming an organ donor, the average effect is close to zero.  Especially given the hypothetical outcome in \citet*{johnson2003}, both compliers and defiers are plausible.  Compliers may think of the default as an endorsement by the experimenter and thus follow it regardless of what it is.  Defiers may think of a default as a threat to their autonomy and thus go against it regardless of what it is.  For this application, the population from which the sample is drawn has clear policy relevance, so we begin by engaging with standard sampling-based results.

Given the plausibility of defiers in this context, a monotonicity assumption might not be reasonable, but there is no obvious alternative, so standard practice is to engage with bounds on defiers, which can be large. As shown at the bottom of Table \ref{tab:auxiliary}, in which we report auxiliary statistics, the smallest possible number of defiers in the sample is 0 because it is always possible for all people who take up to be always takers and all people who do not take up to be never takers, as in the null of Fisher’s exact test.  The largest possible number of defiers in the sample occurs if all $11(=61-50)$ people who do not take up in intervention and all 23 people who take up in control are defiers, yielding a maximum defier share of $30\% ( = (23+11)/115)$.  The estimated Fr\'{e}chet bounds are tighter because they incorporate the estimated marginal distributions of potential outcomes in each arm using the observed takeup rates.  To calculate the upper Fr\'{e}chet bound from (\ref{eq:frechet_bounds}), we consider the bounds on the share of defiers implied by the takeup rates in intervention and control.  The intervention takeup rate implies that the share of defiers in intervention and thus in the sample is at most $18\%(=100\%-82\%)$, and the control takeup rate implies that the share of defiers in control and thus in the sample is at most 43\%, so the share in intervention binds, and the upper Fr\'{e}chet bound is 18\%.  

The Fr\'{e}chet bounds are estimated with error, so there is even greater uncertainty in the number of defiers. The number of defiers in our 95\% smallest credible set ranges from 0 to 34, which represents 0\% to 30\% of the sample of 115. The estimate that minimizes mean squared error lies in the interior of this range, at 9.6\% (Table \ref{tab:mse}).  The 95\% confidence interval on the population fraction of defiers from \citet*{imbens2004}, incorporating the adjustment recommended by \citet*{jun2023} with pretesting at the 0.01 level, also accounts for estimation of the Fr\'{e}chet bounds, so we report it for comparison in Table \ref{tab:auxiliary}.  It is slightly narrower, extending from 0\% to 27\%, but still includes the estimated Fr\'{e}chet bounds. Accounting for estimation of the Fr\'{e}chet bounds is empirically important.  As demonstrated in Table \ref{tab:auxiliary}, if we do not account for estimation of the Fr\'{e}chet bounds and instead construct a 95\% smallest credible set within the estimated Fr\'{e}chet bounds, variation in the likelihood allows us to exclude 9 defiers---a count in the middle of the Fr\'{e}chet bounds translated into whole numbers from 0 to 21. Within the estimated Fr\'{e}chet bounds, the data are almost 20\% more probable under the maximum likelihood estimate than they are under the joint distribution with 9 defiers, illustrating the magnitude of the curvature within the Fr\'{e}chet bounds. 

The maximum likelihood estimates include $21/115=18\%$ defiers, which is equal to the estimated upper Fr\'{e}chet bound.  The maximum likelihood estimates also include $28/115=24\%$ always takers, $66/115=57\%$ compliers, and no never takers. These are the same estimates that would be obtained under an assumption of no never takers and a subsequent deduction of the counts of other types under balance between intervention and control.  Under an assumption of no defiers, the estimated average effect of 39\% is entirely due to compliers, who represent 39\% of the sample. The maximum likelihood estimates preserve the estimated average effect ($39\%=57\% \times (1)+18\% \times (-1)$), but they tell a very different story. Compliers representing 57\% of the sample are affected in one direction and defiers representing 18\% of the sample are affected in the other.  The total share of people affected of $75\%=(57\%+18\%)$ is almost two times as large as the share of people affected under monotonicity. A decision maker who weighs harms to defiers differently from benefits to compliers needs to know the full joint distribution, not merely the average effect. 

In this application, evidence from the design-based model suggests a weaker alternative assumption to monotonicity---a specific joint distribution of potential outcomes that imposes no never takers.  Monotonicity can, of course, still be imposed with acknowledgment that it is a stronger assumption.  While the assumption of no never takers is still strong, by the principle of maximum entropy, it ``assumes the least’’ \citep*{jaynes1968} about the joint distribution of potential outcomes.  From the design-based likelihood in (\ref{eq:likelihood_urn}), the assumption of no never takers can generate the data in at most $(28 \text{ choose } 50-(66-(54-23))) \times (66 \text{ choose } 66-(54-23)) \times (21 \text{ choose } (61-50))=8.5 \times 10^{31}$ randomized assignments representing 3.4\% of the total randomized assignments ($(115 \text{ choose } 61)=2.5 \times 10^{33}$).  The assumption of no defiers can generate the data in at most $(49 \text{ choose } 26) \times (45 \text{ choose } 24) \times (21 \text{ choose } 11)=7.8 \times 10^{31}$ randomized assignments, representing 3.1\% of the total randomized assignments.  An ex post justification for the assumption of no never takers is that though experimental subjects might feel comfortable complying with or defying defaults imposed by the experimenter, they might feel uncomfortable revealing to the experimenter that they would not donate their organs under either default, especially since they do not know when they are asked a hypothetical question if a later question will ask them for a response under the other default.  If an assumption of no never takers is not palatable, evidence for defiers could be used to motivate methods for examining effect heterogeneity, like the collection of additional data or the use of machine learning methods.

If a second stage outcome such as actual organ donation is available, under an assumption of no never takers, the instrumental variable estimate is no longer interpretable as an average treatment effect on compliers without further assumptions.  As discussed by \citet*{angrist1996}, under an additional assumption that average treatment effects are the same for compliers and defiers, the instrumental variable estimate is again interpretable as a LATE. \citet*{dechaisemartin2017} and \citet*{tchetgen2024} propose alternative assumptions to interpret the instrumental variable estimand in the presence of compliers and defiers.

Although the design-based maximum likelihood estimates target the distribution of potential outcomes in the sample rather than the population, an evidence-based assumption of no never takers in the sample can facilitate learning about defiers in the population if data are available on covariates.  Under the assumption of no never takers, it is possible to label some specific people in the sample as compliers and others as defiers: all people in intervention who do not take up must be defiers, and all people who do not take up in control must be compliers.  This reasoning allows us to make claims that the intervention had a causal effect on some specific people in the sample.  Using terminology from \citet*{pearl1999}, under an assumption of no never takers, the intervention was a ``necessary’’ cause of no takeup for all $11(=61-50)$ people assigned intervention who do not take up, and it would have been a ``sufficient’’ cause of takeup for all $31(=54-23)$ people assigned control who do not take up.  Under a monotonicity assumption, it is not possible to directly label any specific people as compliers or defiers because people who take up in intervention can be compliers or always takers, and people who do not take up in control can be compliers or never takers.  An assumption that allows us to label some specific people as compliers and others as defiers could be an important first step to learn their characteristics, improve targeting in future samples, and obtain a larger average effect.

\section{Contributions to the Literature} \label{sec:contributions}

Our primary insight is that a randomized experiment can reveal evidence on an estimand more primitive than the average effect:  the joint distribution of potential outcomes in the sample. The joint distribution is the relevant estimand for decision makers whose utilities depend on the counts of each type, such as the physician in our stylized example.  The physician wants to know how many defiers there are, not merely whether any exist.  An estimate of the joint distribution of potential outcomes can support a monotonicity assumption or inform a specific alternative assumption.  We obtain evidence using the structure of the implemented randomization process to construct a design-based causal model and its resulting likelihood function.  Our work contributes to the literature on design-based econometrics, statistical decision theory, and causal inference without monotonicity.

In his 1935 book on the design of experiments, Fisher notes that the \textit{design} of Charles Darwin’s experiments was “greatly superior” to the “statistical methods available at the time” \citep*{fisher1935}.  In their 2017 handbook chapter on the econometrics of experiments, Susan Athey and Guido Imbens write that they “recommend using statistical methods that are directly justified by randomization, in contrast to the more traditional sampling-based approach that is commonly used in econometrics” \citep*{athey2017}.  The sampling-based approach assumes that people in an experiment are randomly sampled from a population, and it attempts to learn about the population.  The design-based approach, which we adopt, attempts to learn about the sample of people in the experiment, which is useful since participation in experiments and clinical trials can be far from random \citep*{alsan2024} and some samples do not come from a meaningful population \citep*{abadie2020}.

Our use of a design-based model rather than a sampling-based model enables our contributions.  Our ability to learn about defiers with a design-based likelihood may seem paradoxical because it is well known that the traditional sampling-based likelihood does not vary with the number of defiers within the Fr\'{e}chet bounds determined by the estimated marginal distributions of potential outcomes in intervention and control, as we review in \ref{sec:A_sampling}.  Furthermore, it is straightforward to show that the Fr\'{e}chet bounds on defiers always include zero if the estimated average effect shows compliers on average.  A large literature has focused on specifying what we can learn from the nonparametric Fr\'{e}chet bounds in the traditional sampling-based model (\citealt*{balke1997, heckman1997, manski1997mixing, tian2000, zhang2003, imbens2004, fan2010, mullahy2018, li2019, bai2024, semenova2024}).

Our use of the likelihood implied by the design-based model is central to our contributions. The design-based model defines counterfactuals using causal models of potential outcomes from \citet*{neyman1923}, \citet*{welch1937}, \citet*{kempthorne1952}, \citet*{copas1973}, \citet*{rubin1974, rubin1977}, \citet*{greenland1986}, \citet*{holland1986}, and others, and then uses the randomization structure to derive the design-based likelihood. While \citet*{copas1973} derives the design-based likelihood and uses it to test hypotheses about the average effect and the Fisher null hypothesis, our advance is to use the likelihood to learn about a more primitive object: the joint distribution of potential outcomes in the sample, including the number of defiers.  We revisit Copas' design-based model in light of modern causal models, modern assumptions about the absence of defiers, and modern computational advances.  Likelihood-based models can provide exact distributions of experimental data, test statistics, and confidence intervals, but they remain uncommon relative to simulation methods in randomization inference.  One exception is \citet*{ding2019}, who use the design-based likelihood to examine the sensitivity of inference on aggregate parameters to assumptions about the joint distribution of potential outcomes. \citet*{rigdon2015} and \citet*{li2016} construct exact confidence intervals without the design-based likelihood.  We use the design-based likelihood for estimation as well as inference. 

While the design-based model provides evidence about defiers beyond the Fr\'{e}chet bounds in the sample, it is important to consider what evidence, if any, it can provide about the population from which the sample was drawn.  Sometimes, evidence about the population could be more relevant for policy than evidence about the sample.  Under standard sampling assumptions, the experiment does not provide evidence about defiers beyond the Fr\'{e}chet bounds in the population. While standard sampling models assume that the data consist of independent and identically distributed random draws from a population, this assumption is often far from accurate---especially for small, finite populations sampled without replacement. When the population is finite and the sample is drawn without replacement, departure from standard sampling assumptions can  propagate information about defiers in the sample to the population.  Suppose that the 6 people in our stylized example were randomly drawn without replacement from a finite population of 24 people to participate in a pilot study.  As we show in \ref{sec:A_finite}, we can derive an alternative sampling-based likelihood of the joint distribution of potential outcomes in the finite population under sampling without replacement, and it can reveal evidence within the Fr\'{e}chet bounds. In our stylized example, the maximum likelihood estimate for the finite population of 24 based on an experimental sample that was one fourth of the population includes 16 compliers and 8 defiers, exactly four times the estimate in the sample. In terms of likelihood ratios, the evidence revealed within the finite sample is weak, and the evidence revealed within the finite population is even weaker.  Furthermore, as the finite population grows, the evidence weakens, vanishing in the infinite-population limit where there is no evidence within the Fr\'{e}chet bounds---consistent with the standard sampling result.  Nevertheless, weak evidence can improve decision making over no evidence at all.

Our engagement with statistical decision theory in the style of \citet*{wald1949}, rather than classical hypothesis tests, facilitates our progress. Classical hypothesis tests specify a null hypothesis and place the burden of evidence on the alternative hypothesis.  This asymmetric treatment of Type I and Type II errors can imply an unrealistically conservative decision maker \citep*{tetenov2012} and exclude statistical evidence that, while weak, is still informative. To make use of weak evidence from our design-based likelihood, we focus on the statistical decision problem of choosing the correct joint distribution of potential outcomes in the sample, rather than on testing hypotheses about a prespecified distribution under prespecified size and power.  We do not claim to provide a ``test’’ of a monotonicity assumption.  We provide a ``decision’’ about the joint distribution of potential outcomes that can provide evidence for or against a monotonicity assumption.  

Our use of design-based statistical decision theory contributes to the integration of statistical decision theory into econometrics \citep*{manski2004, dehejia2005, manski2007, hirano2008, hirano2009, stoye2012, kitagawa2018, manski2018, manski2019, manski2021, fernandez2024}. Econometrics is a natural discipline to embrace statistical decision theory because statistical decision theory has such a tight relationship with economic theory.  Statistical decision theory is natural to apply in our design-based setting because it does not require large sample assumptions or asymptotic approximations. Our use of design-based statistical decision theory departs from  “asymptotic analysis of statistical decision rules in econometrics” as reviewed in \citet*{hirano2020} in favor of the literature that focuses on statistical decision theory in finite sample settings \citep*{canner1970, manski2007admissible, schlag2007, stoye2007, stoye2009, tetenov2012}.  

We justify our maximum likelihood estimates with justifications commonly employed in statistical decision theory, starting with the justification that our estimates vary systematically with the data, as we show in Figure \ref{fig:largeheatmap}.  One challenge with statistical decision criteria such as minimax and minimax regret is that researchers have shown instances in which they yield rules that do not vary with the data \citep*{schlag2003, manski2004, hirano2009, stoye2009}. In his influential introduction to statistical decision theory, \citet*{ferguson1967} recognizes that because maximum likelihood rules vary with the data, they are “reasonable” because they are better than “just guessing.” “However,” he continues, “decision theory, as developed here, is devoted to the problem of finding optimal rules, so that we do not refer to maximum likelihood estimates [...] unless they turn out naturally to be optimal in some sense.”  

In \ref{sec:A_bayes}, we further justify our maximum likelihood estimates as Bayes optimal under the appropriate conditions, which advances Bayesian decision theory \citep*{chamberlain2011} by incorporating the design-based likelihood.   Bayes optimality implies that our rule is admissible---that is, there is no alternative rule that performs at least as well for all possible joint distributions of potential outcomes, and strictly better for some \citep*{ferguson1967}.  Furthermore, it ensures that our rule cannot be bested in a betting framework \citep*{freedman1969}.  The Bayesian interpretation of our estimates facilitates our auxiliary construction of credible sets on the joint distribution of potential outcomes in Table \ref{tab:auxiliary}, providing an exact Bayesian alternative to classical asymptotic inference on Fr\'{e}chet bounds \citep*{manski1992, horowitz2000, tamer2004, jun2023} and the number of defiers within the Fr\'{e}chet bounds \citep*{imbens2004}. 

In addition to standard decision theoretic justifications, we also offer a perspective that our design-based maximum likelihood estimates are optimal through the principle of maximum entropy \citep*{jaynes1957a, jaynes1957b}, which does not require specification of a prior. \citet*{golan2002} reviews a literature in econometrics that uses the principle of maximum entropy to abstract away from functional form assumptions on the likelihood function.  In our setting, the design of the randomization, which is under the control of the experimenter, determines the functional form of the likelihood function, which also determines the functional form of the entropy. 

Our results provide evidence that can support a monotonicity assumption or a specific alternative, which can be useful because monotonicity assumptions have become widespread, but they are not always realistic. Influential work by \citet*{imbens1994} proposes a monotonicity assumption that assumes away compliers or defiers in the first stage of an instrumental variable model, and \citet*{manski1997monotone} proposes an analogous assumption on treatment response. Monotonicity assumptions are ubiquitous in the analysis of first stages, and they are gaining traction elsewhere \citep*{alsan2025}. First stage monotonicity assumptions are useful for the interpretation of instrumental variable estimates as local average treatment effects on compliers, and we view them as reasonable in many contexts.  However, they can be difficult to defend in some contexts, particularly in health care settings that motivate our work because interventions may be helpful for some but harmful for others.  We interpret our binary outcome as “takeup” in our stylized example and applications so that we can use the familiar terminology of always and never “takers” developed for the first stage \citep*{angrist1996}, but the design-based likelihood is valid for other binary outcomes, like survival.    Given concerns about side effects in medicine, we see potential for useful applications that have a single binary outcome that represents survival.  In those applications, evidence of defiers could motivate researchers to reduce the dosage of a potentially toxic intervention.   

Previous approaches have revealed evidence that violates monotonicity assumptions, but only with additional data beyond our binary intervention and outcome. Machine learning approaches of \citet*{wager2018} and \citet*{semenova2024} require data on covariates to reveal heterogeneous effects. Analysis of side effects in medicine requires data on secondary outcomes (e.g. \citealp{bernard2001}). Specification tests of instrumental variable model assumptions can reveal evidence against monotonicity in the first stage \citep*{imbens1997, richardson2010, huber2012, huber2015testing, kitagawa2015, mourifie2017, machado2019}, and marginal treatment effect methods can reveal evidence against monotonicity in the second stage \citep*{bjorklund1987, heckman1999,  kowalski2023behavior, kowalski2023reconciling}, but such approaches require the additional structure of a two-stage model as well as data on both stages.  Evidence against first stage monotonicity by \citet*{chan2022} requires a multi-valued instrument and the ability to observe a second stage outcome for one value of the first stage outcome.  We contribute to the literature on violations of monotonicity by demonstrating that, even without additional data, our design-based model of an experiment can reveal evidence for or against monotonicity.

We view the development of our visualizations as a secondary contribution.  Figure \ref{fig:six} allows us to depict how the same data can arise from different distributions of potential outcomes within a Fr\'{e}chet set.  Our visualization of how the MLE varies with the data in Figure \ref{fig:largeheatmap} allows us to depict all possible results from an experiment of a given size, demonstrating that we have not engaged in selective reporting.  The visualization allows us to characterize patterns that would be difficult to infer from isolated empirical applications and demonstrates the value of statistical decision theory over testing.

We also offer substantive insights in real empirical applications.  In one, a payment intervention to induce pregnant women to quit smoking, we find no defiers.  In another, a nudge intervention to get more people to hypothetically donate their organs, we find both compliers and the upper Fr\'{e}chet bounds on defiers.  In both applications, the weak evidence that we find can motivate stronger assumptions that researchers can use to learn more.

\section{Implications for Econometric and Applied Research} \label{sec:implications}

We see our demonstration that a design-based model can reveal evidence beyond the average effect with very minimal assumptions and data as a proof-of-concept that could pave the way for the production of stronger evidence that will achieve the ultimate goal of personalizing decisions.  Even if the average effect is the decision-relevant parameter in some contexts, learning about heterogeneity that underlies the average effect is a first step toward targeting that can ultimately yield a larger average effect.  In our second empirical application, under the maximum likelihood joint distribution of potential outcomes, which includes no never takers, we can deduce that the people who are untreated in intervention must be defiers, and the people who are untreated in control must be compliers.  Under monotonicity, it is never possible to deduce that a given individual is a complier, and defiers are assumed away.  Our hope is that the ability to look directly at some compliers and defiers under alternative assumptions motivated by our estimates will pave the way to characterize them, target interventions toward compliers and away from noncompliers, and improve the average effect of future interventions.    

Our findings have already inspired new research from statisticians and econometricians.  Throughout several earlier versions of this paper, we use the design-based likelihood to learn about the joint distribution of potential outcomes \citep*{kowalski2019a,kowalski2019b,christy2024,christy2024a}. In \citet*{christy2024a}, we engage directly with utility functions that depend on the joint distribution of potential outcomes, building on the literature on ``asymmetric counterfactual utilities’’  \citep*{li2019,mueller2023,benmichael2024}, motivated by our work on mammograms, which can help some women but harm others through overdiagnosis \citep*{kowalski2023behavior, kowalski2023reconciling}. Alongside related work on asymmetric counterfactual utilities, our work has prompted engagement from \citet*{koch2025} and inspired \citet*{gelman2025}.  
    
Our work has also inspired \citet*{chen2026}, which explores formal identification and limits to inference in the design-based model.  The authors first confirm that the joint distribution of potential outcomes is formally identified in the design-based model.  They then demonstrate that identification of the joint distribution of potential outcomes permits frequentist and Bayesian inference on a monotonicity hypothesis, but that partial identification in the traditional sampling model nonetheless imposes limits on design-based inference.  Our approaches are distinct and complementary.  First, the target of our inference is the joint distribution of potential outcomes, while these authors target the composite hypothesis of monotonicity.  Second, we demonstrate that the limited information about the joint distribution of potential outcomes can be used for inference in applications, while these authors emphasize a broad set of limits to inference.  In \ref{sec:A_priors}, we engage with the specific result from \citet*{chen2026} most related to our approach, which shows that there exist Bayesian priors over the joint distribution of potential outcomes under which the belief about monotonicity never updates.  We elaborate on this result by showing that the priors they construct assign zero prior weight to many distributions, violating Cromwell's rule \citep*{lindley1985making}. In contrast, our uniform prior is informed by a sampling procedure and assigns non-degenerate weights to all feasible distributions, as we discuss in \ref{sec:A_bayes_cond}.  We also prove that the set of priors that do not update about monotonicity has Lebesgue measure zero and is nowhere dense (that is, topologically ``rare'') in the set of all priors.  We conclude that, while pathological priors exist, they are not generic.

Our work suggests several areas for further methodological research.  The main challenge is to find approaches that strengthen the weak evidence we find on defiers. Design-based likelihoods can be derived for more complex experimental settings, including those with two stages, multi-valued variables, and alternative randomization designs, such as matched pair, stratified, randomized permuted block, and designs that allow for rerandomization.  A promising direction is applying design-based likelihoods to parameters traditionally studied under the lens of partial identification, such as the average treatment effect in experiments with imperfect compliance, missing data, attrition, spillovers, and measurement error.  Another is using design-based likelihoods and entropy to motivate alternative parametric assumptions in auctions \citep*{jun2024} and other designed markets \citep*{abdulkadiroğlu2017}.  Approximation methods and advances in algorithmic efficiency could facilitate calculations in a wide array of samples. Optimal estimators under alternative priors and utility functions could be implemented with straightforward modifications. Estimators could also be developed for alternative estimands, like asymmetric counterfactual utilities, measures of discrimination from audit studies \citep*{kline2021}, and potential outcome quantiles \citep*{cui2023,guggenberger2024}. Finally, while previous results on optimal experimental design focus on estimating the average effect \citep*{bai2022}, other designs, especially those that incorporate additional data, could reveal stronger evidence on the joint distribution of potential outcomes.  

Our design-based maximum likelihood estimate has begun to enter the pedagogy of experimental economics \citep*{list2026}, but a practical impediment to widespread implementation is that applied researchers often do not report enough information about their experiments to replicate them with design-based methods \citep*{young2019}. We use large language models to pull (OpenAI’s GPT-4o-mini) and categorize (OpenAI’s GPT-o3-mini) the randomization processes from 2,080 papers associated with randomized controlled trials from the Abdul Latif Jameel Poverty Action Lab (J-PAL). Only 61\% have enough description for the large language model to confidently categorize the randomization method from the paper or associated AEA RCT Registry entry. However, of those, 78\% use designs captured by design-based likelihoods that we consider. In the full set of papers, over 60\% describe some mechanism through which there could be defiers. Another practical impediment to design-based methods is that instead of reporting absolute numbers, several researchers only report rates or regression results. Mandatory reporting of experimental design and standard sample counts by J-PAL, medical journals, and other authorities could facilitate the application of our design-based likelihood, allowing researchers to learn more from their experiments.

\setcounter{figure}{0}\renewcommand\thefigure{\Alph{subsection}.\arabic{figure}}
\setcounter{table}{0} \renewcommand\thetable{\Alph{subsection}.\arabic{table}}

\renewcommand\thesubsection{Appendix \Alph{subsection}}
\renewcommand\thesubsubsection{\thesubsection .\arabic{subsubsection}}

\titleformat{\section}
  {\bfseries\Large}
  {\thesection}{1em}{}

\titleformat{\subsection}
  {\bfseries\large}
  {\thesubsection}{1em}{\normalfont\large}

\titleformat{\subsubsection}
  {\bfseries}
  {\thesubsubsection}{1em}{\normalfont}

\vspace{5mm}
\section*{Appendix} \label{sec:appendix}
\addcontentsline{toc}{section}{Appendix}

\subsection{Derivations of Likelihoods} \label{sec:A_likelihood}

\subsubsection{Design-Based Likelihoods Can Vary Conditional on the Marginal Distributions of Potential Outcomes}\label{sec:A_likelihood_design}

The design of the experiment---the randomization process---allows us to derive the probability of a realization of the experimental data $\xb$ given the joint distribution of potential outcomes in the sample $\thetab$, which is equal to the likelihood $\mathcal{L}( \thetab \mid \xb )$ by definition. Recall that the distribution of potential outcomes in the sample $\thetab$ consists of the number of always takers $\theta_{11}$, the  number of compliers $\theta_{10}$, the number of defiers $\theta_{01}$, and the number of never takers $\theta_{00}$. We define the joint distribution of potential outcomes in the intervention arm
$\Ib \equiv \big( I_{11}, I_{10}, I_{01}, I_{00} \big)$
as a random vector whose elements represent the numbers of always takers, compliers, defiers, and never takers randomized into intervention; we write its realizations as $\ib = (i_{11}, i_{10}, i_{01}, i_{00})$. Using the randomization design, we can derive the distribution of these unobserved values and use it to compute the distribution of the data $\Xb$.  

Consider an experiment where the randomization design assigns subjects to the intervention or control arm through a series of Bernoulli trials. Then, each subject's probability of assignment to intervention is $p$, and assignments are independent across subjects as well as across principal strata. This independence allows us to write the distribution of $\boldsymbol{I}$ as the product of four independent binomial distributions:
\begin{equation}
\Prob \Big( \Ib = \ib \mid \thetab \Big)
	= \binom{\theta_{11}}{i_{11}} \binom{\theta_{10}}{i_{10}}
		\binom{\theta_{01}}{i_{01}} \binom{\theta_{00}}{i_{00}}p^{\sum_{j,k} i_{j,k}}
		(1-p)^{n - \sum_{j,k} i_{j,k}}. \label{eq:simple_rand}
\end{equation}

Alternatively, in a completely randomized experiment, the randomization design fixes the number of subjects in the intervention arm $m$ and selects any of the possible combinations of $m$ subjects in intervention and $n-m$ subjects in control with equal probability, as though drawing names from a hat. Under this design, $\boldsymbol{I}$ follows a multivariate hypergeometric distribution:
\begin{align}
\Prob \Big( \Ib = \ib \mid \thetab \Big)
= \frac{
	\binom{\theta_{11}}{i_{11}} \binom{\theta_{10}}{i_{10}}
		\binom{\theta_{01}}{i_{01}} \binom{\theta_{00}}{i_{00}} }
	{ \binom{n}{m} }. \label{eq:completely_rand}
\end{align}

While the distribution of potential outcomes in the intervention arm $\Ib$ is unobservable, we can use its data generating process to derive the distribution of the observable data $\Xb$. Note that each subject randomized into the intervention arm with outcome $Y=1$ must be either an always taker or a complier:
$X_{I1} = I_{11} + I_{10}$.
Each subject randomized into the intervention arm with outcome $Y=0$ must be either a never taker or a defier:
$X_{I0} = I_{00} + I_{01}$.
In the control arm, those observed with outcome $Y=1$ must be either always takers that were not randomized into intervention, or defiers that were not randomized into intervention: $X_{C1} = \theta_{11} - I_{11} + \theta_{01} - I_{01}$. And finally, those in the control arm with outcome $Y=0$ must be either never takers that were not randomized into intervention, or compliers that were not randomized into intervention: $X_{C0} = \theta_{00} - I_{00} + \theta_{10} - I_{10}$. Thus, we can write the probability of the observed data $\Xb$ conditional on the joint distribution of potential outcomes $\thetab$ as:
\begin{align*}
\Prob \big( \Xb = \xb \mid \thetab \big)
&= \Prob \Big(
	I_{11} + I_{10} = x_{I1},\
	\big( \theta_{11} - I_{11} \big)
		+ \big( \theta_{01} - I_{01} \big) = x_{C1}, \\*
&\qquad\qquad
	I_{00} + I_{01} = x_{I0},\
	\big( \theta_{00} - I_{00} \big)
		+ \big( \theta_{10} - I_{10} \big) = x_{C0}
	\mid \thetab \Big) \\*
&= \Prob \Big(
	I_{11} + I_{10} = x_{I1},\
	I_{11} + I_{01} = \theta_{11} + \theta_{01} - x_{C1}, \\*
&\qquad\qquad
	I_{00} + I_{01} = x_{I0},\
	I_{00} + I_{10} = \theta_{00} + \theta_{10} - x_{C0}
	\mid \thetab \Big).
\end{align*}

A realization of the data $\xb$ may be produced from more than one realization of the joint distribution of potential outcomes in intervention $\ib$. To find the probability of the realization $\xb$, we sum together the probabilities of each $\ib$ that could have produced it. Using $j$ as the realized number of always takers randomized into intervention $i_{11}$,  we solve the following system of equations for the elements of the joint distribution of potential outcomes in intervention $\Ib$:
\begin{align*}
I_{11} + I_{10} &= x_{I1}, \\
I_{11} + I_{01} &= \theta_{11} + \theta_{01} - x_{C1}, \\
I_{11} + I_{10} + I_{01} + I_{00} &= x_{I1} + x_{I0}, \\
I_{11} &= j.
\end{align*}
Rearranging yields
\begin{align*}
I_{11} &= j \\
I_{10} &= x_{I1} - j \\
I_{01} &= \theta_{11} + \theta_{01} - x_{C1} - j \\
I_{00} &= x_{I0} + x_{C1} + j - \theta_{11} - \theta_{01}.
\end{align*}
The value $j$ is restricted to the set $\mathcal{I}(\xb, \thetab)$ such that $\Ib$ remains within the support implied by $\thetab$, namely
$0 \leq I_{11} \leq \theta_{11}$,
$0 \leq I_{10} \leq \theta_{10}$,
$0 \leq I_{01} \leq \theta_{01}$, and
$0 \leq I_{00} \leq \theta_{00}$.
The probability of a realization of $\Xb$ is just the sum of the probability of each of these realizations of $\Ib$:
\begin{align}
\mathcal{L}( \thetab \mid \xb )
= \Prob \big( \Xb = \xb \mid \thetab \big)
= \sum_{j \in \mathcal{I}(\xb, \thetab)} \Prob \Big(
	I_{11} &= j, \nonumber\\*
	I_{10} &= x_{I1} - j, \nonumber\\*
	I_{01} &= \theta_{11} + \theta_{01}
				- x_{C1} - j, \nonumber\\*
	I_{00} &= x_{I0} + x_{C1} + j - \theta_{11}
				- \theta_{01}
				\mid \thetab \Big).
				\label{eq:likelihood_general}
\end{align}
Equation (\ref{eq:likelihood_general}) represents the general design-based likelihood, where the probability of each distribution of potential outcomes in intervention $\Ib$ given the distribution of potential outcomes in the sample $\thetab$ is informed by the design of the experiment. When subjects are assigned to intervention by a series of Bernoulli trials, $\Ib$ follows the distribution in (\ref{eq:simple_rand}), yielding:
\begin{align}
\mathcal{L}( \thetab \mid \xb )
&= \sum_{j \in \mathcal{I}(\xb, \thetab)} \binom{ \theta_{11} }{ j } \nonumber\\*
	&\qquad \qquad \times
		\binom{ \theta_{10} }{ x_{I1} - j } \nonumber\\*
	&\qquad \qquad \times
		\binom{ \theta_{01} }
			{ \theta_{11} + \theta_{01} - x_{C1} - j } \nonumber\\*
	&\qquad \qquad \times
		\binom{ \theta_{00} }
			{ x_{I0} + x_{C1} + j - \theta_{11}
				- \theta_{01} } 	\nonumber\\*
	&\qquad \qquad \times p^{x_{I1} + x_{I0}} (1-p)^{x_{C1} + x_{C0}} . \label{eq:likelihood_iid}
\end{align}
Alternatively, in a completely randomized experiment, $\Ib$ follows the distribution in (\ref{eq:completely_rand}), yielding:
\begin{align}
\mathcal{L}( \thetab \mid \xb )
&= \sum_{j \in \mathcal{I}(\xb, \thetab)}
	\binom{ \theta_{11} }{ j } \nonumber\\*
	&\qquad \qquad \times
		\binom{ \theta_{10} }{ x_{I1} - j } \nonumber\\*
	&\qquad \qquad \times
		\binom{ \theta_{01} }
			{ \theta_{11} + \theta_{01} - x_{C1} - j } \nonumber\\*
	&\qquad \qquad \times
		\binom{ \theta_{00} }
			{ x_{I0} + x_{C1} + j - \theta_{11}
				- \theta_{01} }
	\bigg/ \binom{n}{m}. \label{eq:likelihood_urn}
\end{align}
In either case, once conditioning on the realized data $\xb$, the likelihood is proportional to
\begin{align}
\mathcal{L}( \thetab \mid \xb )
&\propto \sum_{j \in \mathcal{I}(\xb, \thetab)}
	\binom{ \theta_{11} }{ j } \nonumber\\*
	&\qquad \qquad \times
		\binom{ \theta_{10} }{ x_{I1} - j } \nonumber\\*
	&\qquad \qquad \times
		\binom{ \theta_{01} }
			{ \theta_{11} + \theta_{01} - x_{C1} - j } \nonumber\\*
	&\qquad \qquad \times
		\binom{ \theta_{00} }
			{ x_{I0} + x_{C1} + j - \theta_{11}
				- \theta_{01} }.
\end{align}
This final expression appears in (\ref{eq:likelihood}) and is equivalent to the likelihood in \citet*{copas1973}.  This expression also represents the numerator of the likelihood values we calculate in the stylized example in Section \ref{sec:example}, which vary conditional on the marginal distributions of potential outcomes.

\subsubsection{Canonical Sampling-Based Likelihoods are Flat Conditional on
the Marginal Distributions of Potential Outcomes} \label{sec:A_sampling}

Suppose that subjects are sampled into the experiment with replacement from a population. Let $q_{11}$ represent the share of the population who are always takers, $q_{10}$ the share of compliers, $q_{01}$ the share of defiers, and $q_{00}$ the share of never takers. Sampling with replacement implies that each subject's pair of potential outcomes is independently and identically distributed (IID) according to these shares.

We can derive likelihood functions for the population distribution of potential outcomes $\qb \equiv (q_{11}, q_{10}, q_{01}, q_{00})$ given the data from the experiment, depending on the randomization procedure into intervention. If subjects are assigned to intervention or control via a series of Bernoulli trials, we can view the combination of assignment $Z$ and outcome $Y$ across the $n$ subjects as independent categorical variables whose distribution is defined by the population distribution of potential outcomes $\qb$ and the probability of assignment to intervention $p$:
\begin{align*}
	\Prob\big(Y=1, Z=I \big)
		&= (q_{11} + q_{10}) p \\*
	\Prob\big(Y=0, Z=I \big)
		&= \big( 1 - (q_{11} + q_{10}) \big)p \\*
	\Prob\big(Y=1, Z=C \big)
		&= (q_{11} + q_{01})(1 - p) \\*
	\Prob\big(Y=0, Z=C \big)
		&= \big(1 - (q_{11} + q_{01}) \big)(1 - p).
\end{align*}
The experimental data $\Xb$ are counts of realizations of independent categorical variables.  Therefore, $\Xb$ follows the multinomial distribution with the probabilities above, which yields the likelihood expression:
\begin{align}
\mathcal{L}( \qb \mid \xb )
		&= \Prob(\Xb = \xb \mid \qb) \nonumber\\
		&= \frac{n!}{ x_{I1}! x_{I0}! x_{C1}! x_{C0}! }
			p^{x_{I1} + x_{I0}} (1-p)^{x_{C1} + x_{C0}} \nonumber\\
		&\qquad\times \big( q_{11} + q_{10} \big)^{x_{I1}}
			\big( 1 - (q_{11} + q_{10}) \big)^{x_{I0}} \nonumber\\
		&\qquad\times \big( q_{11} + q_{01} \big)^{x_{C1}}
			\big( 1 - (q_{11} + q_{01}) \big)^{x_{C0}}.
			\label{eq:like_pop_iid}
\end{align}
This likelihood is equivalent to a likelihood appearing in \citet*{barnard1947}.

Alternatively, if the experiment is ``completely randomized,'' such that $m \leq n$ subjects are assigned to intervention by drawing names from a hat and the remainder are assigned to control, we can view the sample as two sets of Bernoulli trials. The first set of $m$ trials represents the subjects sampled into the intervention arm, where there are $X_{I1}$ treated subjects with independent probabilities of being treated equal to the probability that $Y_{I}=1$ in the population, $q_{11} + q_{10}$. The second set of $n-m$ trials represents the subjects sampled into the control arm, where there are $X_{C1}$ treated subjects with independent probabilities of being treated equal to the probability that $Y_{C}=1$ in the population, $q_{11} + q_{01}$. The two sets of Bernoulli trials are independent, allowing us to write the likelihood of a completely randomized experiment as the product of two Binomial random variables $X_{I1}$ and $X_{C1}$:
\begin{align}
\mathcal{L}(\qb \mid \xb)
		&= \binom{m}{x_{I1}} \big( q_{11} + q_{10} \big)^{x_{I1}}
			\big( 1 - (q_{11} + q_{10}) \big)^{m - x_{I1}} \nonumber\\*
		&\qquad\times  \binom{n - m}{x_{C1}} \big( q_{11} + q_{01} \big)^{x_{C1}}
			\big( 1 - (q_{11} + q_{01}) \big)^{n - m - x_{C1}}.
			\label{eq:like_pop_urn}
\end{align}
This likelihood is equivalent to likelihoods from \citet*{barnard1947} and \citet*{kline2021}.

In each of these likelihood functions, the joint distribution of potential outcomes in the population $\qb$ appears only through the transformations $q_{11} + q_{10}$ and $q_{11} + q_{01}$, which are the marginal distributions of the potential outcomes in intervention and control in the population, respectively.  Therefore, the likelihoods of any two distributions in the same population-level Fr\'{e}chet set are equal.  That is, the experimental data contains no information about the joint distribution of potential outcomes in the population beyond the marginal distributions of potential outcomes.  

\subsubsection{Alternative Sampling-Based Likelihoods Can Vary Conditional on the Marginal Distributions of Potential Outcomes} \label{sec:A_finite}

When subjects are sampled from a finite population without replacement, their potential outcomes are no longer independently and identically distributed, and the sampling-based likelihood maintains some of the curvature found in the design-based likelihood.  Let the finite population consist of $\essb \equiv (s_{11},s_{10},s_{01},s_{00})$
where $s_{11}$ denotes the number of always takers, $s_{10}$ the number of compliers,
$s_{01}$ the number of defiers, and $s_{00}$ the number of never takers.
Let the total population size be $k=s_{11}+s_{10}+s_{01}+s_{00}$. Suppose that $n$ subjects are sampled without replacement from this population.
As before, let $\thetab=(\theta_{11},\theta_{10},\theta_{01},\theta_{00})$ denote the counts of each type in the experimental sample, where $\theta_{11}+\theta_{10}+\theta_{01}+\theta_{00}=n$.  Under sampling without replacement, the distribution of $\thetab$ conditional on the
population counts $\essb$ follows the multivariate hypergeometric distribution:

$$
\mathbb{P}(\thetab \mid \essb)=\frac{
\binom{s_{11}}{\theta_{11}}
\binom{s_{10}}{\theta_{10}}
\binom{s_{01}}{\theta_{01}}
\binom{s_{00}}{\theta_{00}}
}{
\binom{k}{n}
}.
$$

Among the sampled subjects, the experiment randomly assigns subjects to intervention and control. Because the experimental sample itself is random, the likelihood of the population counts $\essb$ must integrate over all possible realizations of the sample type counts $\thetab$. Applying the law of total probability yields
$$
\mathcal{L}(\essb\mid \xb) = \sum_{\thetab\in\Thetab} \mathcal{L}(\thetab\mid \xb)\mathbb{P}(\thetab\mid \essb).
$$
Note that, conditional on the sample distribution $\thetab$, the experimental outcome $\xb$ does not depend on the entire population distribution $\essb$. Substituting the expression for $\mathbb{P}(\thetab\mid \essb)$ from above gives
$$
\mathcal{L}(\essb\mid \xb)=\sum_{\thetab\in\Thetab}
\mathcal{L}(\thetab\mid \xb)
\frac{
\binom{s_{11}}{\theta_{11}}
\binom{s_{10}}{\theta_{10}}
\binom{s_{01}}{\theta_{01}}
\binom{s_{00}}{\theta_{00}}
}{
\binom{k}{n}
},
$$
where $\mathcal{L}(\thetab\mid \xb)$ is the design-based likelihood of the experiment sample, which is given by equation (\ref{eq:likelihood_iid}) under simple randomization or equation (\ref{eq:likelihood_urn}) under complete randomization, both of which can be found in \ref{sec:A_likelihood_design}. 

Unlike the population likelihoods under sampling with replacement we derive in \ref{sec:A_sampling}, which depend on the joint distribution of potential outcomes only through the marginal distributions, the population likelihood under sampling without replacement is a function of the entire joint distribution of potential outcomes.  As a result, the population likelihood under sampling without replacement varies within the population Fr\'{e}chet bounds on defiers: if 6 people are drawn from a finite population of 24 without replacement, 2 of 3 randomized into intervention take up, and 1 of 3 randomized into control take up, then the finite population consisting of 8 always takers, 8 compliers, and 8 never takers has a likelihood of 0.0374, while the finite population consisting of 16 compliers and 8 defiers has a likelihood of 0.0548, despite the fact that these joint distributions have the same marginal distributions.

\subsection{Justifications for Our Maximum Likelihood Decision Rule} \label{sec:justifications}

\subsubsection{Preliminaries on Statistical Decision Theory} \label{sec:A_bayes_prelim}

Maximum likelihood estimates are ubiquitous in statistics and econometrics, and we offer three justifications for their use in our design-based model of an experiment.  The first two justifications are common in statistical decision theory, which treats estimation and inference as decision problems under uncertainty and looks for optimal estimators---or ``decision rules.''  Suppose a decision maker wishes to guess the joint distribution of potential outcomes in the sample. We write the decision maker's guess as $\widehat{\thetab}$, and we define a utility function $u$ over a guess $\widehat{\thetab}$ and the true distribution $\thetab$. We also allow the decision maker to choose a randomized guess, which ascribes a probability distribution over the possible values of $\thetab$. We define the decision maker's utility over a randomized guess $\rho$ as the expected utility of guessing according to the probabilities ascribed by $\rho$:
\begin{align}
	U(\rho, \thetab) = \sum_{\widehat{\thetab} \in \Thetab} u(\widehat{\thetab}, \thetab) \rho(\widehat{\thetab}).
\end{align}

The decision maker chooses a decision rule that maps the observable data into (possibly) randomized guesses.\footnote{We conflate here the standard definitions of ``randomized decision rules'' and ``behavioral decision rules'' \citep*{ferguson1967} for expositional clarity. In settings of perfect recall, such as the setting we study here, the space of randomized and behavioral decision rules is equivalent \citep*{kuhn1953}.} We write such a rule as $f: \boldsymbol{\mathcal{X}} \to \Delta$, where $\boldsymbol{\mathcal{X}}$ is the space of possible data realizations, $\Thetab$ is the space of possible distributions of potential outcomes, and $\Delta$ is the space of distributions over $\Thetab$. Given a true distribution of potential outcomes $\thetab$, the decision maker's expected utility from following a decision rule $f$ is the expected value of $U(f(\Xb), \thetab)$ with respect to the experimental outcome $\Xb$:
\begin{align*}
	EU(f ,\thetab)
		&= \mathbb{E} \big[U \big(f(\Xb), \thetab\big)
							\mid \thetab \big] 				
		= \sum_{\xb \in \boldsymbol{\mathcal{X}}}
			\ \sum_{\widehat{\thetab} \in \Thetab}
			u(\widehat{\thetab}, \thetab)
			\mathcal{L}(\thetab \mid \xb)
			f(\xb)(\widehat{\thetab}) 
\end{align*}

One sense in which a decision rule can be optimal is in terms of Bayes expected utility, which is the average expected utility obtained according to some subjective prior distribution over $\thetab$:
\begin{align}
	EU_{\pi}(f) = \mathbb{E}_{\pi}\big[EU(f, \thetab)\big]
		= \sum_{\thetab \in \Thetab} EU(f, \thetab) \pi(\thetab), \label{eq:EU}
\end{align}
where $\pi \in \Delta$ is a subjective prior belief about $\thetab$. Decision rules that maximize this criterion are said to be ``Bayes optimal.''

To define the maximum likelihood decision rule, let $\widehat{\Thetab}(\xb)$ be the set of joint distributions $\thetab$ that maximize the likelihood given the observed data $\xb$:
\begin{align*}
	\widehat{\Thetab}(\xb) = {\arg \max}_{\thetab} \mathcal{L}(\thetab \mid \xb)
\end{align*}
There are finitely many vectors of four integers that sum to the number of participants in the experiment $n$, so $\widehat{\Thetab}(\xb)$ is nonempty.  We define the maximum likelihood decision rule as an estimator $\widehat{\thetab}_{MLE}$ that places equal weight on each maximizer of the likelihood:
\begin{align*}
	\mathbb{P} (\widehat{\thetab}_{MLE}(\xb) = \thetab \mid \xb) =
	\begin{cases}
		\dfrac{1}{\#\{\widehat{\Thetab}(\xb)\}}, & \text{if} \quad \thetab \in \widehat{\Thetab}(\xb),  \\
		0, &\text{otherwise,}
	\end{cases}
\end{align*}
where $\#\{\bullet\}$ is the counting measure.  When the likelihood is unimodal, the decision rule chooses the maximizer with probability one.

\subsubsection{Our Maximum Likelihood Decision Rule Varies Systematically with the Data} \label{sec:vary}

\begin{figure}[hbtp!]
	\caption{Visualization of Our Proposed Statistical Decision Rule:  How the MLE Varies with All Possible Data from Samples of 50 and 200 with Half in Intervention} 
	\includegraphics[width=0.90\linewidth]{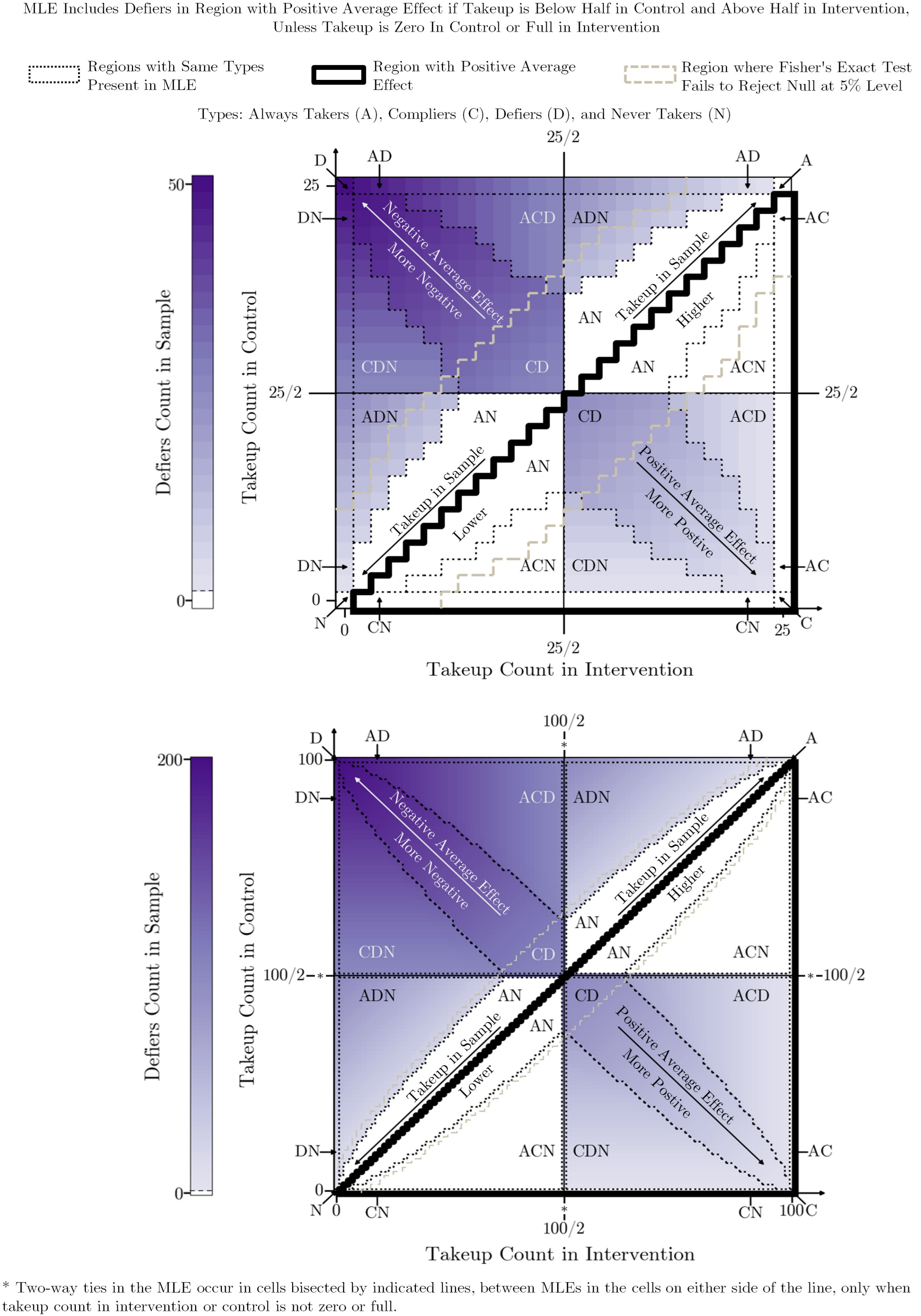}
	\label{fig:largeheatmap}
\end{figure}

Our first justification for our maximum likelihood decision rule is that it varies with the data, so it is ``reasonable'' as argued by \citet*{ferguson1967} in his introduction to statistical decision theory.   Direct inspection of the likelihood in (\ref{eq:likelihood}) does not reveal obvious patterns in the maximum likelihood estimate.  However, by computing the maximum likelihood estimate for every possible realization of the data in fixed sample sizes, we can observe clear patterns in the estimates.  Figure \ref{fig:largeheatmap} illustrates how the maximum likelihood estimates from our proposed decision rule vary systematically with all possible realizations of the data in completely randomized experiments of 50 and 200 subjects with  half in intervention. In a sample of 200 with half in intervention, there are $10,201=101\times101$ possible realizations of the data and $1,373,301={203\choose 3}$ possible joint distributions of potential outcomes, so the visualization reflects results of grid searches over $14,009,043,501= 10,201\times1,373,301$ total elements. 

In the subfigure for each sample size, the horizontal axis gives the takeup count in intervention, and the vertical axis gives the takeup count in control.  Each cell represents a possible realization of the data, which specifies an estimated Fr\'{e}chet set in terms of takeup count in intervention and control.  The visualization demonstrates that an estimated Fr\'{e}chet set can also be specified with the estimated average effect, which increases diagonally downward, and the total takeup count in the sample, which increases diagonally upward.  An estimated Fr\'{e}chet set can also be specified with an estimated average effect and a takeup rate in control, as is common in empirical practice.

The figure provides a visualization of our statistical decision rule by reporting statistics from our maximum likelihood estimates of the numbers of always takers, compliers, defiers, and never takers in the sample. Maximum likelihood estimates in cells within dotted outlines include the same types: always takers (A), compliers (C), defiers (D), and never takers (N).  Darker purple shading within cells indicates higher counts of defiers in the maximum likelihood estimates. 

The figure shows clear patterns in how the MLEs vary with the data, which are visibly similar when we increase the sample size by a factor of four. MLEs in the corners only include one type: always takers if takeup is full in intervention and control, compliers if takeup is full in intervention and zero in control, defiers if takeup is zero in intervention and full in control, and never takers if takeup is zero in intervention and control.  MLEs along the edges and diagonals only include the two types in the corners they connect.  

In the convex lens-shaped region around the upward-sloping diagonal, MLEs only include always and never takers: the two types in the null hypothesis of the \citet*{fisher1935} exact test.  In samples of 50 and 200, the region in which Fisher’s exact test fails to reject the null hypothesis at the 5\% level is depicted within the grey dashed line.  It is larger than the region in which the MLE includes only always and never takers in both sample sizes.  

The figure illustrates how our design-based maximum likelihood decision rule reveals evidence that is obscured by Fisher’s exact test.  Fisher’s exact test only considers the null hypothesis that the joint distribution of potential outcomes includes only always and never takers. If it rejects, it does not suggest an alternative.  If it fails to reject, a joint distribution of potential outcomes that includes types other than always and never takers can maximize the likelihood.  With our use of statistical decision theory, we can harness the richer evidence conveyed in the figure beyond the grey dashed lines.    

In the convex lens-shaped region around the downward sloping diagonal, MLEs only include compliers and defiers.  This region is related to the region in which an exact test of the null hypothesis that everyone is affected does not reject.  There are some realizations of the data near the center of the region in which Fisher’s exact test fails to reject the null hypothesis that no one is affected, but our maximum likelihood estimates indicate that everyone is affected.  These estimates are relevant to empirical practice for researchers who wonder if they obtained a near-zero estimated average effect not because no one was affected but because everyone was affected.  

Sometimes, the MLE can be a tie between more than one distribution.  MLEs are unique in the sample of 50 with half in intervention, but two-way ties occur in the sample of 200.  In cells bisected by the lines indicated with asterisks, the midlines in which takeup is exactly half in intervention or control, ties occur between MLEs in cells on either side of the line.  Only at the edges of the figure do the ties result in a unique MLE because the MLE is the same on either side of the line. For experiments with intervention and control arms unequal in size, ties appear in less predictable patterns.

The MLEs include three types in all other regions, with two regions for each combination.  Even though the maximizer of the likelihood only includes two types in our stylized example from Figure \ref{fig:six}, maximization of the likelihood sometimes requires trading off between the higher likelihood of fewer types and the higher likelihood of more balance between intervention and control within each type.  If restricting the sample to two types requires too much imbalance between intervention and control, the maximizer of the likelihood includes three types.\footnote{Viewing Pascal’s triangle as a representation of $N$ choose $K$ can help us to visualize the tradeoff between fewer types (moving to the next row of the triangle, increasing $N$ within each type) and more balance within each type (moving to the middle of the triangle within a row, moving $K$ closer to $N/2$ within each type). We thank Elizabeth Ananat for this point.}  There are no regions in which a distribution that maximizes the likelihood contains all four types.  Therefore, many joint distributions of potential outcomes never maximize the likelihood for any realization of the data. For example, if a sample of 50 consisted of 18 always takers, 22 compliers, 8 defiers, and 2 never takers, the most probable realization of the data would be takeup of 20  in intervention and 13  in control, but the corresponding MLE estimate of 26 always takers, 14 compliers, 0 defiers, and 10 never takers would miss the 8 defiers entirely. We have observed MLEs that contain all four types in experiments with intervention and control arms unequal in size, although we have only observed them tied with MLEs that contain three types.  

The region outlined in bold, in which the estimated average effect is positive, is of particular interest because MLEs in the shaded portion of the region include both defiers and compliers, contrary to monotonicity assumptions, as in the stylized example in Figure \ref{fig:six}.  In all even-sized samples up to 200, the MLE includes defiers when takeup is below half in control and above half in intervention, unless takeup is zero in control or full in intervention.  The experiment depicted in Figure \ref{fig:six} falls into the region with defiers because the takeup rate is 1/3 in control and 2/3 in intervention. As shown, the MLE always includes defiers when the estimated average effect is greater than half, unless takeup is zero in control or full in intervention, and it can include defiers even when Fisher’s exact test rejects at the 5\% level. This pattern might seem surprising or even disconcerting because at the edges of the shaded regions, small changes in the data can yield large changes in the estimated numbers of defiers. However, while point estimates can change abruptly, credible sets on either side of a transition overlap substantially, reflecting the underlying continuity of the likelihood. We recommend that applied researchers report credible sets alongside maximum likelihood estimates, as we do in our applications. Moreover, the persistence of the pattern in samples of 50 and 200 suggests that the region in which the MLE includes defiers will persist in larger samples.

The shading demonstrates that the MLE reveals evidence beyond the average effect: holding the estimated average effect constant along upward sloping diagonal lines, MLEs can include zero or many defiers. \citet*{copas1973} claims that the likelihood is always maximized at a joint distribution of potential outcomes that preserves the direct estimates of the marginal distributions of potential outcomes from equation (\ref{eq:marginal_est}) and which has either the maximal or minimal number of defiers.  In practice, we often find that the likelihood is maximized at a distribution that deviates slightly from these marginal estimates.

\subsubsection{Our Maximum Likelihood Decision Rule is Bayes Optimal} \label{sec:A_bayes}

\paragraph{We Motivate Conditions for Bayes Optimality:}\label{sec:A_bayes_cond}

Our second justification for our proposed maximum likelihood decision rule is that it is Bayes optimal under a specific utility function and a specific prior. Bayes optimal decision rules can be derived for alternative utility functions and priors through the numerical solution of a linear programming problem, but we offer motivations for the specifications we have chosen, acknowledging their subjectivity. For the utility function, we specify a discrete utility function that assigns zero utility for incorrect guesses and an arbitrary amount of strictly positive utility for correct guesses:
\begin{align}
    u(\widehat{\thetab},\thetab) = \mathbf{1}_{\{\widehat{\thetab}=\thetab\}}. \label{eq:util}
\end{align}
For the prior, we specify a prior that is uniform over all possible distributions of potential outcomes for $n$ subjects:
\begin{align}
    \pi(\thetab)=\frac{1}{\#\{\Thetab\}},\ \forall\thetab \in \Thetab. \label{eq:prior}
\end{align}

One motivation for our utility function  is that we can interpret it as the payoff function in a game that rewards ``being correct,'' as in the well-known Monty Hall game, where the contestant receives a prize only if they guess the correct door.  As we review below, the maximum likelihood decision rule under this utility function is the maximum \emph{a posteriori} (MAP) rule, which chooses the mode of the posterior distribution.  Under a uniform prior, the posterior distribution is proportional to the likelihood, and the MAP estimate equals the MLE estimate.

The desirable feature of our uniform prior is that it assigns equal weight to all distributions in the same Fr\'{e}chet set, so it is ``uninformative'' in the sense that it does not confer any information about the joint distribution of potential outcomes beyond their marginal distributions.  This uninformative prior is consistent with the beliefs of a statistician who places equal weight on all values of defiers within the estimated Fr\'{e}chet bounds.  Furthermore, we can view our uniform prior as the result of sampling from an underlying population over which we hold an uninformative hyperprior.  In sampling settings, it is common to assume a Dirichlet hyperprior over population shares.  We choose a Dirichlet hyperprior over the population shares of always takers, compliers, defiers, and never takers with all concentration parameters set to one, which is equivalent to an uninformative uniform distribution for the population.  Under standard assumptions on sampling from the population, the potential outcomes of subjects are IID, and the sample distribution of potential outcomes follows a multinomial law conditional on the population distribution.  Combining the Dirichlet hyperprior and the sampling process yields a Dirichlet-multinomial prior over the counts of each type in the sample.  The Dirichlet-multinomial prior with concentration parameters set to one is equivalent to our uniform prior.   In our stylized example in Figure \ref{fig:six}, our uniform prior implies that the prior is the same in all three rows, so the updated posterior is proportional to the likelihood.    

Even for other prior specifications, the ratio of the beliefs between any two distributions updates as long as the likelihood ratio differs from one, including within Fr\'{e}chet sets.  For example,  consider an alternative prior that is uniform over each type \textit{for each person} in the sample.  This prior arises from a highly informative degenerate hyperprior that assumes the population shares of each type are equal.  The resulting prior over sample distributions varies within the sample Fr\'{e}chet sets, and is therefore also informative.  In our stylized example in Figure \ref{fig:six}, the resulting prior is 0.0220 for the first row, 0.0293 for the second row, and 0.0037 for the third row.  The updated posterior distribution maintains curvature within the Fr\'{e}chet set and the ratio of the prior to the posterior is highest for the third row that contains two defiers.  To achieve a flat posterior distribution across the rows depicted in Figure \ref{fig:six}, the prior would have to be very specific and informative: the prior for the first row with no defiers would have to be exactly 0.75 (=(20/8)/(20/6)) times as large as the second row with one defier and 1.5 (=(20/8)/(20/12)) times as large as the prior for the third row with two defiers.

\paragraph{A Standard Proof Establishes Bayes Optimality:} \label{sec:A_bayes_proof}

Proposition \ref{prop:optim} establishes the Bayes optimality of the maximum likelihood decision rule under the conditions described above.\footnote{We thank Andriy Norets and Thomas Wiemann for bringing the standard conditions that establish Bayes optimality to our attention.} 

\begin{proposition}\label{prop:optim}
    When utility $u$ takes the form in (\ref{eq:util}) and the prior belief $\pi$ takes the form in (\ref{eq:prior}), the maximum likelihood decision rule $\widehat{\thetab}_{MLE}$ is Bayes optimal:
    \begin{align*}
        \widehat{\thetab}_{MLE} \in {\arg \max}_f EU_\pi (f).
    \end{align*}
\end{proposition}

\begin{proof}
We first establish that the maximum \emph{a posteriori} decision rule is Bayes optimal for our chosen loss function given any prior. The maximum \emph{a posteriori} decision rule $f^*_\pi$ selects the maxima of the posterior distribution of $\thetab$.  Formally, let $\widehat{\Thetab}_{\pi}(\xb)$ be the set of  $\thetab$ values that maximize the posterior distribution given the observed data $\xb$, i.e.
\begin{align}
	\widehat{\Thetab}_{\pi}(\xb)
		&= \argmax{\thetab \in \Thetab} \Prob \big( \thetab \mid \xb \big) \nonumber \\*
		&= \argmax{\thetab \in \Thetab} \mathcal{L}(\thetab \mid \xb) \pi(\thetab), \label{eq:map_set}
\end{align}
where $\pi \in \Delta$ is the prior belief about $\thetab$.  The decision rule $f^*_\pi$ can then be defined as follows:
\begin{align}
	f^*_\pi(\xb)(\thetab) = \begin{cases}
		\dfrac{1}{\#\{\widehat{\Thetab}_{\pi}(\xb)\}}	
			& \text{if}\ \thetab \in \widehat{\Thetab}_{\pi}(\xb), \\
		0 									
			& \text{o.w.}
	\end{cases} \label{eq:f_opt}
\end{align}

Now, let $g$ be an arbitrary decision function.  The Bayes expected utility for decision function $g$ is
\begin{align*}
	\E\big[ EU(g, \thetab) \big] 
		&= \sum_{\thetab \in \Thetab} EU(g, \thetab) \pi(\thetab)
		= \sum_{\thetab \in \Thetab} 	
				\Bigg[ \sum_{\xb \in \boldsymbol{\mathcal{X}}}
				\mathcal{L} (\thetab \mid \xb) 
				g(\xb)(\thetab)\Bigg]
				\pi (\thetab). 
\end{align*}
Note that, under our choice of utility function, the expected utility of decision rule $g$ is the probability that $g$ correctly guesses the distribution of potential outcomes $\thetab$. By rearranging terms in the summation, we can bound the Bayes expected utility of $g$:
\begin{align*}
	\mathbb{E}\big[EU(g, \thetab)\big]
		&= \sum_{\xb \in \boldsymbol{\mathcal{X}}}
			\sum_{\thetab \in \Thetab}
			\bigg(
				\mathcal{L}(\thetab \mid \xb) 
				g(\xb)(\thetab) \pi(\thetab)\bigg) \\*				
		&\leq \sum_{\xb \in \boldsymbol{\mathcal{X}}}
			\ \sum_{\thetab \in \Thetab}
				\bigg( g(\xb)(\thetab)
					\max_{\thetab^\prime \in \Thetab} 
						\Big\{ \mathcal{L}(\thetab^\prime \mid \xb) 
							\pi(\thetab^\prime) \Big\} \bigg) \\*
		&= \sum_{\xb \in \boldsymbol{\mathcal{X}}}
			\Bigg[ \max_{\thetab^\prime \in \Thetab} 
				\Big\{ \mathcal{L}(\thetab^\prime \mid \xb)
					\pi(\thetab^\prime) \Big\}
				\underbrace{ \bigg( \sum_{\thetab \in \Thetab}
					g(\xb)(\thetab) \bigg)}_{=1} \Bigg].
\end{align*}
This bound is precisely the Bayes expected utility achieved by decision rule $f^*_\pi$:
\begin{align*}
	\E\big[EU(f^*_\pi, \thetab)\big] 
		&= \sum_{\xb \in \boldsymbol{\mathcal{X}}}
			\sum_{\thetab \in \Theta} 
				\bigg( \mathcal{L}(\thetab \mid \xb)
				f^*_\pi(\xb)(\thetab) \pi(\thetab) \bigg) \\*
		&=  \sum_{\xb \in \boldsymbol{\mathcal{X}}}
			\sum_{\thetab \in \widehat{\Thetab}_{\pi}(\xb)} 
				\bigg( \frac{1}{\#\{\widehat{\Thetab}_{\pi}(\xb)\}}
					\max_{\thetab^\prime \in \Thetab}
					\Big\{ \mathcal{L}(\thetab^\prime \mid \xb) 
						\pi(\thetab^\prime) \Big\} \bigg) \\*
		&= \sum_{\xb \in \boldsymbol{\mathcal{X}}}
			\Bigg[ \max_{\thetab^\prime \in \Thetab}
				\Big\{ \mathcal{L}(\thetab^\prime \mid \xb) 
					\pi(\thetab^\prime) \Big\}
				\underbrace{ \bigg( \sum_{\thetab \in \widehat{\Thetab}_{\pi}(\xb)}
					\frac{1}{\#\{\widehat{\Thetab}_{\pi}(\xb)\}} \bigg) }_{=1} \Bigg].				
\end{align*}
Thus, since $f^*_\pi$ achieves the upper bound on the Bayes expected utility of any decision rule, we conclude that $f^*_\pi$ is Bayes optimal.  Finally, observe that when the prior distribution $\pi(\thetab)$ is constant, the maximizers of the posterior distribution in (\ref{eq:map_set}) are simply the maximizers of the likelihood, and the Bayes rule $f^*_\pi$ in (\ref{eq:f_opt}) simplifies to our maximum likelihood decision rule.
\end{proof}

\paragraph{Other Bayes Optimal Estimators Can Be Derived, Including a Minimum Mean Squared Error Estimator:} \label{sec:A_bayes_mse}

For alternative utility functions or priors, the optimal rule can be derived using the design-based likelihood.  As an example, we derive the optimal estimator for another common utility function, negative squared error:
\begin{align}
    u(\widehat{\thetab}, \thetab) = - \|\widehat{\thetab} - \thetab\|^2,\label{eq:mse}
\end{align}
where $\|\bullet\|$ is the Euclidean norm.  For simplicity, we restrict our attention to nonrandomized decision rules, and we do not require the estimates to be integer valued.  In this setting, Proposition \ref{prop:MSE} establishes the optimal estimator with squared error utility under any prior:
\begin{proposition}\label{prop:MSE}
    When utility $u$ takes the form in (\ref{eq:mse}), the optimal nonrandomized decision rule, which we refer to as the minimum mean squared error rule, selects the posterior mean:
    \begin{align*}
        \widehat{\thetab}_{MSE}(\xb) 
                = \mathbb{E}\big[\thetab\mid\xb\big]
    \end{align*}
\end{proposition}
\begin{proof}
    We will first show that the posterior mean maximizes the expectation of utility given $\xb$:
    \begin{align}
        \max_{\widehat{\thetab} \in \Thetab} 
            \mathbb{E}\Big[
                -\|\widehat{\thetab} - \thetab\|^2 
                \mid \xb
            \Big]. \label{eq:mse_max}
    \end{align}
    Expanding the objective function yields:
    \begin{align*}
        \mathbb{E} \Big[
            -\|\widehat{\thetab}-\thetab\|^2 \mid \xb 
        \Big]
        &= -\mathbb{E}\Big[ 
            \Big\|
                \big(
                    \widehat{\thetab} - \mathbb{E}[\thetab \mid \xb]
                \big) + 
                \big(
                    \mathbb{E}[\thetab \mid \xb] - \widehat{\thetab}
                \big)
            \Big\|^2 \mid \xb
        \Big] \\
        &= - \mathbb{E}\Big[
            \big\| \widehat{\thetab} - \mathbb{E}[\thetab \mid \xb] \big\|^2
            + 2\big( \widehat{\thetab} - \mathbb{E}[\thetab \mid \xb] \big)^\top
                \big( \mathbb{E}[\thetab \mid \xb] - \thetab \big)  \\
        &\qquad \qquad
            + \big\| \mathbb{E}[\thetab \mid \xb] - \thetab \big\|^2
            \mid \xb 
            \Big] \\
        &= - \big\|
                \widehat{\thetab} - \mathbb{E}[\thetab \mid \xb]
            \big\|^2
            - \mathbb{E} \Big[
                \big\|
                     \mathbb{E}[\thetab \mid \xb] - \thetab
                \big\|^2 \mid \xb
            \Big].
    \end{align*}
    The second term does not depend on $\widehat{\thetab}$, so the objective is maximized by setting $\widehat{\thetab} = \mathbb{E}[\thetab \mid \xb]$.

    Now, let $\pi(\xb)$ be the probability of the data realization $\xb$ under the prior $\pi$, and consider the Bayes expected utility of an arbitrary nonrandomized decision rule $\widehat{\thetab}$:
    \begin{align*}
        \mathbb{E}\big[EU(\widehat{\thetab},\thetab)\big]
        &= \sum_{\thetab \in \Thetab} \sum_{\xb \in \mathcal{X}}
            -\big\| \widehat{\thetab}(\xb) - \thetab \big\|^2 
            \mathcal{L}(\thetab \mid \xb)
            \pi(\thetab) \\
        &= \sum_{\xb \in \mathcal{X}} 
            \mathbb{E} \Big[
                -\big\| \widehat{\thetab}(\xb) - \thetab \big\|^2 \mid x
            \Big] \pi(\xb) \\
        &\leq \sum_{\xb \in \mathcal{X}} 
            \mathbb{E} \Big[
                - \big\|
                    \widehat{\thetab}_{MSE}(\xb) - \thetab
                \big \|^2 \ \mid \xb 
            \Big] \pi(\xb) \\
        &= \mathbb{E}\Big[
            EU\big(
                \widehat{\thetab}_{MSE}, \thetab
            \big)
        \Big ].
    \end{align*}
The inequality follows from the fact that $\mathbb{E}[\thetab \mid \xb]$ solves (\ref{eq:mse_max}).  Therefore, we conclude that the minimum mean squared error decision rule $\widehat{\thetab}_{MSE}$ is Bayes optimal.    
\end{proof}

For a given utility function and prior, the optimal rule may not take a convenient form such as the MLE or the posterior mean, but it can still be computed by solving the linear programming problem of maximizing Bayes expected utility in (\ref{eq:EU}) with respect to the (finitely-many) values of $f(\xb)(\thetab)$, such that $\sum_\thetab f(\xb)(\thetab) = 1$ for each $\xb$.

\paragraph{The Relative Performance of the Optimal Rule Increases in Even Sample Sizes from 2 to 200:}\label{sec:A_bayes_perform}

The Bayes optimality of our proposed maximum likelihood decision rule allows us to quantify the gains from using the optimal rule over alternatives.  We consider two alternative decision rules.  The first alternative decision rule, which we refer to as the ``Fr\'{e}chet rule,’’ estimates the marginal distributions of potential outcomes directly according to (\ref{eq:marginal_est}), and then chooses randomly among the joint distributions of potential outcomes in the corresponding Fr\'{e}chet set.  The Fr\'{e}chet rule $\widehat{\thetab}_F$ satisfies:
\begin{align*}
	\mathbb{P}(\widehat{\thetab}_{F}(\xb) = \thetab) =
	\begin{cases}
		\dfrac{1}{\#\mathcal{F}(\hat{\theta}_{1 \bullet}(\xb), \hat{\theta}_{\bullet 1}(\xb); n)}
		& \text{if}\; \thetab \in \mathcal{F}(\hat{\theta}_{1 \bullet}(\xb), \hat{\theta}_{\bullet 1}(\xb); n),  \\
		0 & \text{otherwise}.
	\end{cases}
\end{align*}
This decision rule captures the behavior of a statistician who assumes that the experimental data are uninformative within Fr\'{e}chet sets and is therefore indifferent between any two distributions within the same Fr\'{e}chet set.

The second alternative decision rule, which we refer to as the ``monotonicity rule,'' approximates the behavior of a statistician who uses the monotonicity assumption of \citet*{imbens1994} in the population to estimate the  share of compliers or defiers in the sample.  The rule uses the observed difference in takeup rate  between the intervention and control arms $\delta$ to approximate the share of compliers or defiers. In particular, if the difference is non-negative, the rule imposes a no-defiers monotonicity restriction, and if the difference is non-positive, it imposes a no-compliers monotonicity restriction.

Under this decision rule, the inferred number of compliers or defiers is given by 
$$
\hat\theta_{10}=n\cdot \max\{\delta, 0\},\quad \hat\theta_{01}=n\cdot\max\{-\delta,0\},
$$
rounded to the nearest whole number. Note that when intervention and control arms are the same size, these expressions are always whole numbers. The remaining types, always takers and never takers, are inferred by assuming balance of type shares across both arms. In the case where there are compliers and no defiers, every treated individual in control must be an always taker, and because assignment is randomized, the proportion of always takers in control should be equal to the proportion of always takers in the sample, giving the estimate $\hat\theta_{11}=nx_{C1}/(x_{C1}+x_{C0})$.  The remaining individuals must all be never takers. Conversely, if there are defiers and no compliers, every treated individual in intervention must be an always taker, so the estimate of always takers is $\hat\theta_{11}=nx_{I1}/(x_{I1}+x_{I0})$ and the remaining individuals are never takers.  In summary, the monotonicity rule $\widehat{\thetab}_M$ chooses
\begin{align*}
    \widehat{\thetab}_M(\xb) 
    &= \big(\hat{\theta}_{11},\hat{\theta}_{10},\hat{\theta}_{01},\hat{\theta}_{00}\big) \\
    &= \bigg(\mathbf{1}_{\{\delta > 0\}}\left[n\frac{x_{C1}}{x_{C1}+x_{C0}}\right] + \mathbf{1}_{\{\delta\leq0\}}\left[n\frac{x_{I1}}{x_{I1}+x_{I0}}\right],\ n\cdot \max\{\delta,0\},\ n \cdot \max\{-\delta,0\}, \\ 
    &\qquad n-\hat{\theta}_{11}-\hat{\theta}_{01}-\hat{\theta}_{10}\bigg)
\end{align*}
with probability one.

We calculate the exact Bayes expected utility achieved by our optimal maximum likelihood rule and each of the two alternative rules.  For a given sample size, we enumerate each of the finitely many feasible sample distributions of potential outcomes $\thetab$.  For each distribution $\thetab$, we compute the expected utility of our decision rules by comparing the distribution of estimates they produce under each data realization $\xb$ to the true value of $\thetab$.  We average the resulting utilities over the values of $\xb$ according to the likelihood.   We then average these expected utilities according to a uniform prior for $\thetab$ to get the Bayes expected utility.

\begin{figure}[hbt!]
	\caption{Under Optimality Conditions, For Increasing Sample Sizes, Performance of Proposed Maximum Likelihood Rule Increases Relative to Fr\'{e}chet and Monotonicity
 Rules}  
	\centering
	\includegraphics[width=0.7\linewidth]{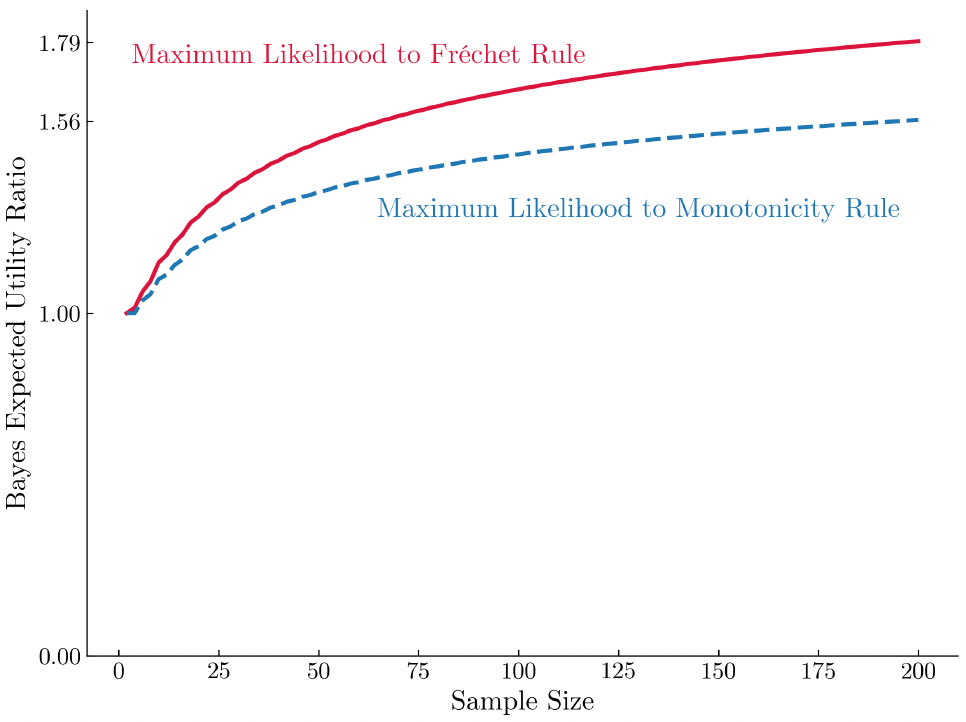}
	\label{fig:bayes}
\end{figure}

Figure \ref{fig:bayes} plots the ratio of the Bayes expected utility achieved by our proposed maximum likelihood rule to the Bayes expected utility achieved by each of the two alternative rules for even sample sizes between 2 and 200.  The gains from using the optimal rule over either of the alternatives increase with the sample size.  In a sample size of 200, the Bayes expected utility of the proposed maximum likelihood rule is 1.79 times higher than that of the Fr\'{e}chet rule, and 1.56 times higher than that of the monotonicity rule.

\paragraph{Conditions for Bayes Optimality Enable the Construction of Credible Sets:}\label{sec:A_credible}

Under the conditions in which our proposed maximum likelihood decision rule is Bayes optimal, we can quantify uncertainty in estimation through Bayesian credible sets.  We construct credible sets on the joint distribution of potential outcomes in the sample using the posterior belief implied by a uniform prior and the design-based likelihood.  A credible set of level $\alpha$ is a set of joint distributions of potential outcomes covering a posterior probability mass of at least $\alpha$.  We report the smallest credible set, which is the credible set containing the fewest points.  We construct these sets with the following algorithm:
\begin{samepage}
\begin{enumerate}
	\item Compute the posterior probability of every joint distribution of potential outcomes $\thetab$ given the observed data $\xb$.
	\item Order the joint distributions from highest to lowest posterior probability.
	\item Compute the cumulative probability mass according to this ordering.
	\item Starting with the highest probability distribution (which is the MLE), add joint distributions of potential outcomes to the credible set until the probability mass of the credible set exceeds $\alpha$.  In the case of ties, we add all distributions with the same probability simultaneously.
\end{enumerate}
\end{samepage}
We also report the range of values for each type---always takers, compliers, defiers, and never takers---present in the credible set constructed on the joint distribution.  Note that the ranges reported on each type are not independent.  For example, the maximal values of the ranges for always takers, compliers, defiers, and never takers may not constitute an element of the credible set on the joint distribution of potential outcomes.

\subsubsection{Our Maximum Likelihood Decision Rule Maximizes Entropy} \label{sec:entropy}

We can also motivate our maximum likelihood rule through the principle of maximum entropy \citep*{jaynes1957a, jaynes1957b}, which unites the theory of information with statistical physics. In Section \ref{sec:example}, we discuss how differences in entropy drive variation in the likelihood such that the maximum likelihood distribution is also the maximum entropy distribution. Formally, the entropy is proportional to the logarithm of the number of assignments.  Because the logarithm is monotonic, the distribution that maximizes the number of assignments also maximizes entropy.  \citet*{jaynes1968} advocates for the application of the principle of maximum entropy to circumvent the subjective specification of a prior.  To make Bayesian decision-making more objective, one option is to choose the least informative prior. Jaynes’ alternative is to choose the least informative updated distribution. The distribution that can generate the data in the greatest number of ways---the distribution that maximizes entropy and our design-based likelihood---is the least informative updated distribution.  In the Monty Hall game, a celebrated application of Bayesian decision making, it is possible to make the same decision without specifying a Bayesian prior using a maximum likelihood decision rule and the principle of maximum entropy \citep*{wang2016}.

\subsection{Applications: Auxiliary Statistics}

\begin{figure}[H]
	\centering
	\caption{Standard Statistics, Proposed Design-Based Maximum Likelihood Estimates, and Auxiliary Statistics}
    \makebox[\textwidth][c]{
    \hspace{-0.00\textwidth}
	\includegraphics[width=1.0\linewidth]{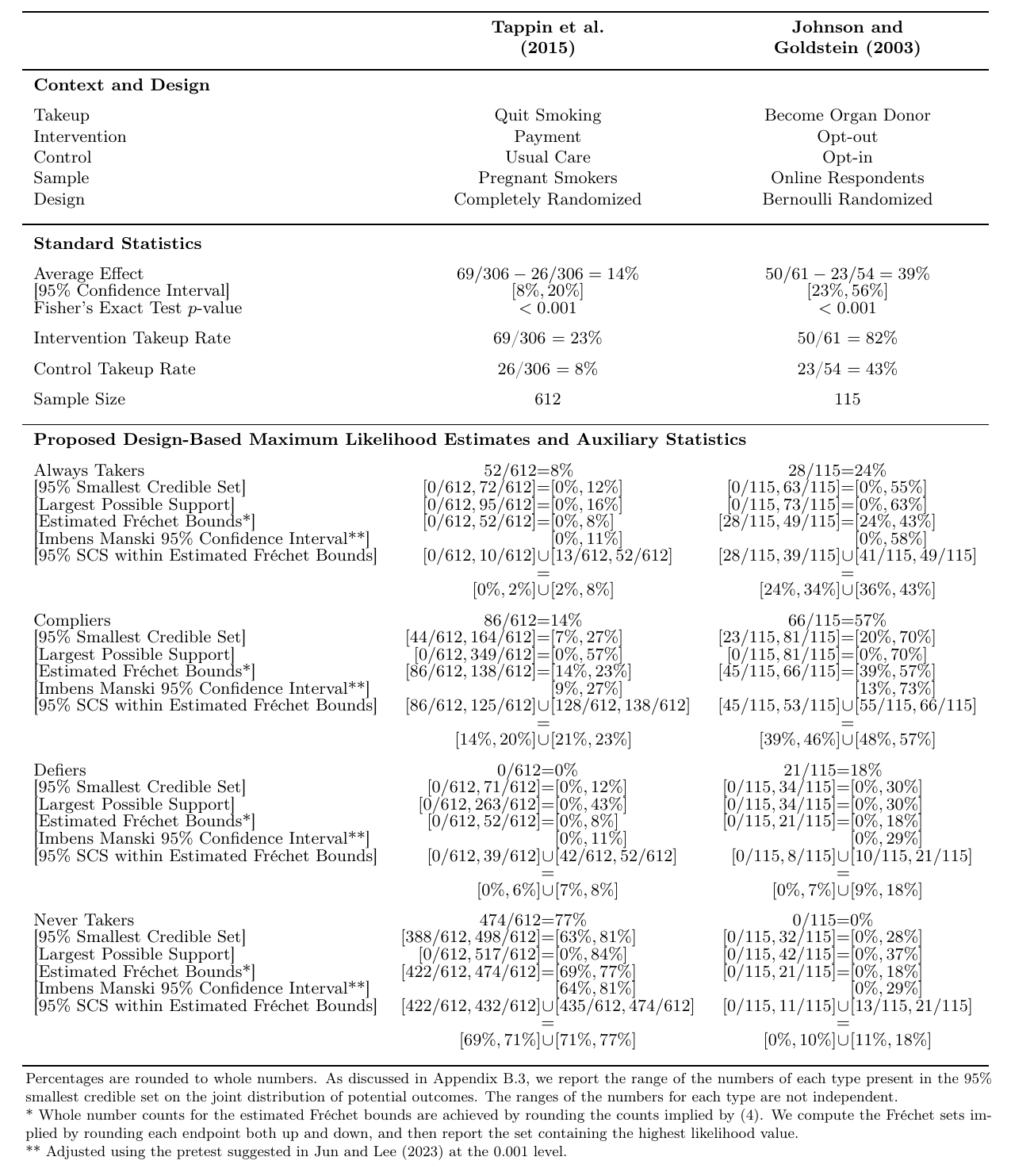}
    }
	\label{tab:auxiliary}
\end{figure}

\begin{table}[H]
    \centering
    \caption{Standard Statistics and Proposed Design-Based \\ Minimum Mean Squared Error Estimates}
	\includegraphics[width=\linewidth]{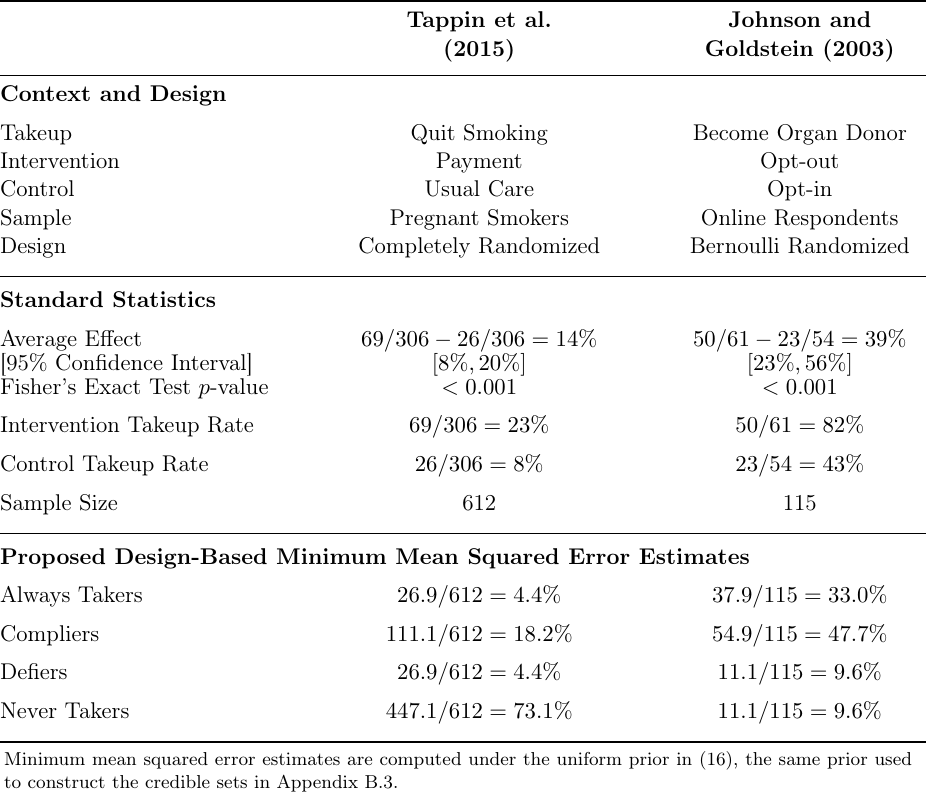}
	\label{tab:mse}
\end{table}

\subsection[Responses to a Follow-up Paper by Chen, Roth, and Spiess]{Responses to a Follow-up paper by \citet*{chen2026}} \label{sec:A_priors}

\subsubsection{Priors That Do Not Update About Monotonicity Can Have Undesirable Features}

Throughout several earlier versions of this paper, we use the design-based likelihood to learn about the joint distribution of potential outcomes \citep*{kowalski2019a,kowalski2019b,christy2024,christy2024a}. Recently, \cite*{chen2026} (CRS, hereafter) have confirmed the formal identification of the joint distribution of potential outcomes in the design-based model, and they have emphasized limits to frequentist and Bayesian inference on a monotonicity assumption that there are either compliers or defiers, but not both.  Their analysis is distinct and complementary to ours.  The result most closely related to our work shows, through a constructive proof, the existence of priors over the possible joint distributions of potential outcomes in the sample $\thetab$ under which the belief in a monotonicity hypothesis does not update for any realization of the data (Proposition 5.1).  We expand on this result by further characterizing the priors that they construct in their proof, as well as the entire space of priors that do not update with respect to monotonicity.  

Using our stylized example from Section \ref{sec:example}, we demonstrate that the priors constructed in the proof of Proposition 5.1 of CRS can have undesirable features.  First, the priors $\pi$ for a sample of 6 people constructed in the fashion of CRS that do not update about monotonicity  have limited support:
\begin{align*}
    \thetab \in \text{supp } \pi \Rightarrow 
    \thetab = \big(\theta_{11}, \theta_{10}, \theta_{01}, \theta_{00}\big)\in 
    \Big\{& (6, 0, 0, 0), (4, 0, 0, 2), (2, 0, 0, 4), (0, 0, 0, 6), \\
    & (0, 1, 5, 0), (0, 3, 3, 0), (0, 5, 1, 0) \Big\}.
\end{align*}
There are ${6+4-1\choose 4-1}=84$ ways to allocate 6 people into 4 types, but these priors place weight on only 7---a violation of Cromwell's rule, which advises against prior beliefs of 0 or 1 on any event \citep*{lindley1985making}. Furthermore, these 7 consist only of distributions with at most two types present, so these priors place zero weight on any distribution with three or four types.  The construction of non-updating priors in CRS relies on combinations of priors whose supports contain distributions with only two types present, so all priors constructed in this manner violate Cromwell's rule.  Moreover, under such priors, the possible data realizations are only those in which an even number of subjects take up, so data with odd total takeup---like the data we consider in our stylized example, where 2 subjects take up in intervention and 1 takes up in control---are impossible under these priors. The prior thus assigns our stylized example zero probability \emph{a priori}.

\subsubsection{Priors that Do Not Update About Monotonicity Are Rare}

The undesirable features of the priors offered in the constructive proof of CRS  motivate a characterization of priors that do not update about monotonicity.  We consider the space of all priors that do not update about monotonicity, not just those constructed by CRS.  We show that the set of priors that do not update about monotonicity has Lebesgue-measure zero and is nowhere dense in the set of all possible priors.

We first establish some additional notation. Fix a finite sample size $n\geq 3$. Let $\Thetab$ be the set of joint distributions of types for sample size $n$. Define $\Thetab_0 \equiv \{\thetab \in \Thetab : \theta_{01} = 0\ \text{or}\ \theta_{10} = 0\}$ as the set of monotonic distributions and define $\Thetab_1$ as the set of non-monotonic distributions---the complement of $\Thetab_0$. Let $\Delta$ be the set of priors on $\Thetab$:
\begin{align*}
    \Delta \equiv \bigg\{\pi:\Thetab\to [0,1]: \sum_{\thetab \in \Thetab} \pi(\thetab)=1\bigg\}.
\end{align*}
Since $\Thetab$ is a finite subset of $\mathbb{Z}_+^4$, the set of priors $\Delta$ is a subset of a finite-dimensional Euclidean space. Let $\mathcal{X}_\pi$ be the support of data realizations given a prior $\pi$, and define the set of non-updating priors $\mathcal{N}$:
\begin{align*}
\mathcal{N} \equiv \big\{\pi\in \Delta:\pi(\Thetab_0\mid \xb)=\pi(\Thetab_0) \,\forall \xb\in\mathcal{X}_\pi\big\}.
\end{align*}
Proposition \ref{prop:rare} presents our claim. We thank our excellent research assistant Shuheng Zhang for the formulation and proof of this proposition.
\begin{proposition}\label{prop:rare}
    Let the set of priors $\Delta$ be a subset of some finite-dimensional Euclidean space $\mathbb{R}^m$ equipped with the standard Euclidean topology and Lebesgue measure.  Then, the set of non-updating priors $\mathcal{N}\subseteq \Delta$ has Lebesgue measure zero and is nowhere dense (``rare'') in $\Delta$.
\end{proposition}

\begin{proof}
    
For data realization $\xb$, define functions $A_\xb, B_\xb : \Delta \to \mathbb{R}$ as:
\begin{align*}
    A_\xb(\pi) \equiv \sum_{\thetab\in \Thetab_0}\pi(\thetab)\mathcal{L}(\thetab\mid \xb),
    \qquad 
    B_\xb(\pi) \equiv \sum_{\thetab\in \Thetab_1}\pi(\thetab)\mathcal{L}(\thetab\mid \xb).
\end{align*}  
Define the function $F_\xb:\Delta \to \mathbb R$ as:
\begin{align*}
    F_\xb(\pi) = A_\xb(\pi)\pi(\Thetab_1)-B_\xb(\pi)\pi(\Thetab_0).
\end{align*}
The posterior belief in monotonicity when $\xb\in\mathcal{X}_\pi$ is given by $\pi(\Thetab_0\mid \xb)=A_\xb(\pi)(A_\xb(\pi)+B_\xb(\pi))^{-1}.$
A prior does not update its belief on monotonicity given a data realization $\xb$ if and only if $A_\xb(\pi)=\pi(\Thetab_0)(A_\xb(\pi)+B_\xb(\pi))$ by Bayes' rule, or equivalently, if $F_\xb(\pi)=0$.  Note that $A_\xb(\pi)$, $B_\xb(\pi)$, $\pi(\Thetab_0)$, and $\pi(\Thetab_1)$ are all linear functions of the coordinates $\{\pi(\thetab)\}_{\thetab\in \Thetab}$. Thus, $F_\xb(\pi)$ is a quadratic polynomial in these coordinates. Now, for any given $\xb$, define $Z_\xb$ as the set of all priors that do not update their belief of monotonicity given $\xb$:
\begin{align*}
Z_\xb \equiv \big\{\pi\in\Delta:\pi(\Thetab_0\mid \xb)=\pi(\Thetab_0)\big\} = \big\{\pi\in \Delta:F_\xb(\pi)=0 \big\}.
\end{align*}
The set $Z_\xb$ is the intersection of $\Delta$ with the roots of the degree 2 polynomial $F_\xb(\cdot)$. 

Now, we will show there exists some $\xb^*\in \mathcal{X}$ such that $F_{\xb^*}(\cdot)$ is not the zero polynomial. Suppose for the sake of contradiction that $F_\xb(\cdot)=0$ for all $\xb\in\mathcal{X}$. Choose an $\xb$, $\thetab_0\in\Thetab_0$, and $\thetab_1\in \Thetab_1$. Consider the family of priors $\pi_t$ of the form $\pi_t(\thetab_0)=t$ and $\pi_t(\thetab_1)=1-t$ for 
$t\in(0,1)$, so $\pi_t(\thetab)=0$ for all other $\thetab$. For such a $\pi_t$, we have that  $A_\xb(\pi_t)=t\mathcal{L}(\thetab_0\mid \xb)$, $ B_\xb(\pi_t)=(1-t)\mathcal{L}(\thetab_1\mid \xb)$, $\pi_t(\Thetab_0)=t$, and $\pi_t(\Thetab_1)=1-t$. Thus,
\begin{align*}
F_\xb(\pi_t) 
&= t\mathcal{L} (\thetab_0\mid \xb)(1-t)-(1-t)\mathcal{L}(\thetab_1\mid \xb)t\\
&= t(1-t)\big(\mathcal{L}(\thetab_0\mid \xb)-\mathcal{L}(\thetab_1\mid \xb)\big).
\end{align*}
Since $F_\xb(\cdot)=0$, we have that $F_\xb(\pi_t)=0$ for all $t\in(0,1)$, which requires $\mathcal{L}(\thetab_0\mid \xb)=\mathcal{L}(\thetab_1\mid \xb)$. But $\xb$, $\thetab_0$, and $\thetab_1$ were arbitrarily chosen, which would imply that $\mathcal{L}(\thetab_0\mid \xb)=\mathcal{L}(\thetab_1\mid \xb)$ for all $\xb\in \mathcal{X}$, $\thetab_0\in \Thetab_0$, and $\thetab_1\in\Thetab_1$, which contradicts our main likelihood curvature result. Thus, there exists at least one $\xb^*\in \mathcal{X}$ such that $F_{\xb^*}(\cdot)$ is not the zero function.

Fixing such an $\xb^*$, we have that $F_{\xb^*}(\pi)$ is polynomial and not the zero polynomial, so its roots form a set of Lebesgue measure zero. Thus, its intersection with $\Delta$, that is, the set $Z_{\xb^*}$, also has Lebesgue measure zero. Since 
$$
\mathcal{N}=\bigcap_{\xb\in\mathcal{X}}Z_\xb\subset Z_{\xb^*},
$$
it also must be measure zero. Then the set of priors that do not update on monotonicity is a set of Lebesgue measure zero. Furthermore, the interior of the closure of $Z_{\xb^*}$ in $\Delta$ is empty, so $\mathcal{N}$ is nowhere dense in $\Delta$.  
\end{proof}

Proposition \ref{prop:rare} implies that most Bayesians will learn about monotonicity, since the set of priors that do not update is measure zero.  And, since this class of priors is nowhere dense in the space of priors, there are priors that cannot be closely approximated by non-updating priors.    While CRS demonstrate the theoretical existence of priors that do not update about monotonicity, we show that these priors are rare.

\singlespacing
\bibliographystyle{chicago}
\bibliography{why}

@article{abadie2002,
	title={Bootstrap tests for distributional treatment effects in instrumental variable models},
	author={Abadie, Alberto},
	journal={Journal of the American Statistical Association},
	volume={97},
	number={457},
	pages={284--292},
	year={2002},
	publisher={Taylor \& Francis}
}

@article{abadie2003,
	title={Semiparametric instrumental variable estimation of treatment response models},
	author={Abadie, Alberto},
	journal={Journal of econometrics},
	volume={113},
	number={2},
	pages={231--263},
	year={2003},
	publisher={Elsevier}
}

@article{abadie2020,
  title={Sampling-based versus design-based uncertainty in regression analysis},
  author={Abadie, Alberto and Athey, Susan and Imbens, Guido W and Wooldridge, Jeffrey M},
  journal={Econometrica},
  volume={88},
  number={1},
  pages={265--296},
  year={2020},
  publisher={Wiley Online Library}
}

@article{abdulkadiroğlu2017,
author = {Abdulkadiroğlu, Atila and Angrist, Joshua D. and Narita, Yusuke and Pathak, Parag A.},
title = {Research Design Meets Market Design: Using Centralized Assignment for Impact Evaluation},
journal = {Econometrica},
volume = {85},
number = {5},
pages = {1373-1432},
doi = {https://doi.org/10.3982/ECTA13925},
url = {https://onlinelibrary.wiley.com/doi/abs/10.3982/ECTA13925},
year = {2017}
}

@techreport{alsan2025,
	 title = "Mean Reversion in Randomized Controlled Trials: Implications for Program Targeting and Heterogeneous Treatment Effects",
	 author = "Alsan, Marcella and Cawley, John and Doyle, Joseph J, Jr. and Skelley, Nicholas",
	 institution = "National Bureau of Economic Research",
	 type = "Working Paper",
	 series = "Working Paper Series",
	 number = "33369",
	 year = "2025",
	 month = "January",
	 doi = {10.3386/w33369},
	 URL = "http://www.nber.org/papers/w33369"
}

@article{alsan2024,
  title={Representation and extrapolation: evidence from clinical trials},
  author={Alsan, Marcella and Durvasula, Maya and Gupta, Harsh and Schwartzstein, Joshua and Williams, Heidi},
  journal={The quarterly journal of economics},
  volume={139},
  number={1},
  pages={575--635},
  year={2024},
  publisher={Oxford University Press}
}

@incollection{athey2017,
  title={The econometrics of randomized experiments},
  author={Athey, Susan and Imbens, Guido W},
  booktitle={Handbook of Economic Field Experiments},
  volume={1},
  pages={73--140},
  year={2017},
  publisher={Elsevier}
}

@article{angrist1996,
  title={Identification of causal effects using instrumental variables},
  author={Angrist, Joshua D and Imbens, Guido W and Rubin, Donald B},
  journal={Journal of the American Statistical Association},
  volume={91},
  number={434},
  pages={444--455},
  year={1996},
  publisher={Taylor \& Francis}
}

@article{bai2022,
  title={Optimality of matched-pair designs in randomized controlled trials},
  author={Bai, Yuehao},
  journal={American Economic Review},
  volume={112},
  number={12},
  pages={3911--3940},
  year={2022},
  publisher={American Economic Association 2014 Broadway, Suite 305, Nashville, TN 37203}
}

@article{bai2024,
  title={On the Identifying Power of Monotonicity for Average Treatment Effects},
  author={Bai, Yuehao and Huang, Shunzhuang and Moon, Sarah and Shaikh, Azeem M and Vytlacil, Edward J},
  journal={arXiv preprint arxiv:2405.14104},
  year={2024}
}

@article{balke1997,
  title={Bounds on treatment effects from studies with imperfect compliance},
  author={Balke, Alexander and Pearl, Judea},
  journal={Journal of the American Statistical Association},
  volume={92},
  number={439},
  pages={1171--1176},
  year={1997},
  publisher={Taylor \& Francis}
}

@article{barnard1947,
    author = {Barnard, G. A.},
    title = "{Significance Tests for 2 x 2 Tables}",
    journal = {Biometrika},
    volume = {34},
    number = {1-2},
    pages = {123-138},
    year = {1947},
    issn = {0006-3444},
    doi = {10.1093/biomet/34.1-2.123},
    url = {https://doi.org/10.1093/biomet/34.1-2.123},
    eprint = {https://academic.oup.com/biomet/article-pdf/34/1-2/123/552350/34-1-2-123.pdf},
}

@article{benmichael2024,
  title={Policy learning with asymmetric counterfactual utilities},
  author={Ben-Michael, Eli and Imai, Kosuke and Jiang, Zhichao},
  journal={Journal of the American Statistical Association},
  pages={1--14},
  year={2024},
  publisher={Taylor \& Francis}
}

@article{bernard2001,
author = {Bernard, Gordon R. and Vincent, Jean-Louis and Laterre, Pierre-Francois and LaRosa, Steven P. and Dhainaut, Jean-Francois and Lopez-Rodriguez, Angel and Steingrub, Jay S. and Garber, Gary E. and Helterbrand, Jeffrey D. and Ely, E. Wesley and Fisher, Charles J.},
title = {Efficacy and Safety of Recombinant Human Activated Protein C for Severe Sepsis},
journal = {New England Journal of Medicine},
volume = {344},
number = {10},
pages = {699-709},
year = {2001},
doi = {10.1056/NEJM200103083441001},
    note ={PMID: 11236773},

URL = { 
        https://doi.org/10.1056/NEJM200103083441001
    
},
eprint = { 
        https://doi.org/10.1056/NEJM200103083441001
}    
}

@incollection{boole1854, 
	place={London}, 
	title={Of Statistical Conditions}, 
	booktitle={An Investigation of the Laws of Thought: On Which Are Founded the Mathematical Theories of Logic and Probabilities}, 
	publisher={Walton and Maberly},
	author={Boole, George}, 
	year={1854}, 
	pages={295–319},
	chapter=19 
}

@article{bjorklund1987,
  title={The estimation of wage gains and welfare gains in self-selection models},
  author={Bj{\"o}rklund, Anders and Moffitt, Robert},
  journal={The Review of Economics and Statistics},
  pages={42--49},
  year={1987},
  publisher={JSTOR}
}

@article{canner1970,
  title={Selecting one of two treatments when the responses are dichotomous},
  author={Canner, Paul L},
  journal={Journal of the American Statistical Association},
  volume={65},
  number={329},
  pages={293--306},
  year={1970},
  publisher={Taylor \& Francis}
}

@incollection{chamberlain2011,
    author = {Chamberlain, Gary},
    isbn = {9780199559084},
    title = { Bayesian Aspects of Treatment Choice},
    booktitle = {The Oxford Handbook of Bayesian Econometrics},
    publisher = {Oxford University Press},
    year = {2011},
    month = {09},
    doi = {10.1093/oxfordhb/9780199559084.013.0002},
    url = {https://doi.org/10.1093/oxfordhb/9780199559084.013.0002},
    eprint = {https://academic.oup.com/book/0/chapter/334661705/chapter-ag-pdf/44445461/book_38595_section_334661705.ag.pdf},
    pages = {10-39}
}

@article{chan2022,
  title={Selection with variation in diagnostic skill: Evidence from radiologists},
  author={Chan, David C and Gentzkow, Matthew and Yu, Chuan},
  journal={The Quarterly Journal of Economics},
  volume={137},
  number={2},
  pages={729--783},
  year={2022},
  publisher={Oxford University Press}
}

@misc{chen2026,
      title={Testing Monotonicity in a Finite Population}, 
      author={Jiafeng Chen and Jonathan Roth and Jann Spiess},
      year={2026},
      eprint={2512.25032},
      archivePrefix={arXiv},
      primaryClass={econ.EM},
      url={https://arxiv.org/abs/2512.25032}, 
}

@article{christy2024a,
      title={Starting Small: Prioritizing Safety over Efficacy in Randomized Experiments Using the Exact Finite Sample Likelihood}, 
      author={Neil Christy and A. E. Kowalski},
      year={2024},
      journal={arXiv preprint arxiv:2407.18206}
}

@article{christy2024,
      title={Counting Defiers in Health Care with a Design-Based Likelihood for the Joint Distribution of Potential Outcomes}, 
      author={Neil Christy and Amanda Ellen Kowalski},
      year={2024},
      journal={arXiv preprint arXiv:2412.16352d}
}

@misc{christy2025dbmle,
  author       = {Christy, Neil and Kowalski, Amanda and Zhang, Shuheng},
  title        = {dbmle: Design-Based Maximum Likelihood Estimation for Always Takers, Compliers, Defiers, and Never Takers},
  year         = {2025},
  howpublished = {\url{https://pypi.org/project/dbmle/}},
  note         = {Python package version 0.0.2}
}

@article{copas1973,
    author = {Copas, J. B.},
    title = "{Randomization models for the matched and unmatched 2 x 2 tables}",
    journal = {Biometrika},
    volume = {60},
    number = {3},
    pages = {467-476},
    year = {1973},
    issn = {0006-3444},
    doi = {10.1093/biomet/60.3.467},
    url = {https://doi.org/10.1093/biomet/60.3.467},
    eprint = {http://oup.prod.sis.lan/biomet/article-pdf/60/3/467/576103/60-3-467.pdf},
}

@book{cox1958,
	title = {Planning of Experiments},
	year = {1958},
	author = {D. R. Cox},
	publisher = {New York, NY: Wiley}
}

@article{cui2023,
  title={Policy learning with distributional welfare},
  author={Cui, Yifan and Han, Sukjin},
  journal={arXiv preprint arXiv:2311.15878},
  year={2023}
}

@article{dawid2022,
  title={Effects of causes and causes of effects},
  author={Dawid, A Philip and Musio, Monica},
  journal={Annual Review of Statistics and Its Application},
  volume={9},
  number={1},
  pages={261--287},
  year={2022},
  publisher={Annual Reviews}
}

@article{dechaisemartin2017,
  title={Tolerating defiance? Local average treatment effects without monotonicity},
  author={De Chaisemartin, Clement},
  journal={Quantitative Economics},
  volume={8},
  number={2},
  pages={367--396},
  year={2017},
  publisher={Wiley Online Library}
}

@article{dehejia2005,
  title={Program evaluation as a decision problem},
  author={Dehejia, Rajeev H},
  journal={Journal of Econometrics},
  volume={125},
  number={1-2},
  pages={141--173},
  year={2005},
  publisher={Elsevier}
}

@article{ding2019,
  title={Model-free causal inference of binary experimental data},
  author={Ding, Peng and Miratrix, Luke W},
  journal={Scandinavian Journal of Statistics},
  volume={46},
  number={1},
  pages={200--214},
  year={2019},
  publisher={Wiley Online Library}
}

@article{fan2010,
 ISSN = {02664666, 14694360},
 URL = {http://www.jstor.org/stable/40664510},
 author = {Yanqin Fan and Sang Soo Park},
 journal = {Econometric Theory},
 number = {3},
 pages = {931--951},
 publisher = {Cambridge University Press},
 title = {SHARP BOUNDS ON THE DISTRIBUTION OF TREATMENT EFFECTS AND THEIR STATISTICAL INFERENCE},
 volume = {26},
 year = {2010}
}

@book{ferguson1967, 
	place={New York}, 
	title={Mathematical Statistics: A Decision Theoretic Approach}, 
	publisher={Academic Press}, 
	author={Ferguson, Thomas S.}, 
	year={1967}
}

@article{fernandez2024,
  title={Robust Bayes Treatment Choice with Partial Identification},
  author={Fern{\'a}ndez, Andr{\'e}s Aradillas and Montiel Olea, Jos{\'e} Luis and Qiu, Chen and Stoye, J{\"o}rg and Tinda, Serdil},
  journal={arXiv preprint arXiv:2408.11621},
  year={2024}
}

@BOOK{ferrie2017,
  title     = "Statistical physics for babies",
  author    = "Ferrie, Chris",
  publisher = "Sourcebooks",
  series    = "Baby university",
  month     =  dec,
  year      =  2017,
  address   = "Naperville, IL",
  language  = "en"
}

@book{fisher1935,
  title={Design of Experiments},
  author={Fisher, R.A.},
  year={1935},
  publisher={Edinburgh: Oliver and Boyd},
  edition={1st}
}

@article{frangakis2002,
author = {Frangakis, Constantine E. and Rubin, Donald B.},
title = {Principal Stratification in Causal Inference},
journal = {Biometrics},
volume = {58},
number = {1},
pages = {21-29},
doi = {10.1111/j.0006-341X.2002.00021.x},
year = {2002}
}

@article{frechet1957,
 ISSN = {03731138},
 URL = {http://www.jstor.org/stable/1401672},
 author = {M. Fr\'echet},
 journal = {Revue de l'Institut International de Statistique / Review of the International Statistical Institute},
 number = {1/3},
 pages = {23--40},
 publisher = {[International Statistical Institute (ISI), Wiley]},
 title = {Les tableaux de corrélation et les programmes linéaires},
 volume = {25},
 year = {1957}
}

@article{freedman1969,
  title={Bayes' method for bookies},
  author={Freedman, David A and Purves, Roger A},
  journal={The Annals of Mathematical Statistics},
  volume={40},
  number={4},
  pages={1177--1186},
  year={1969},
  publisher={JSTOR}
}

@article{golan2002,
  title={Information and entropy econometrics—editor's view},
  author={Golan, Amos},
  journal={Journal of econometrics},
  volume={107},
  number={1-2},
  pages={1--15},
  year={2002},
  publisher={Elsevier}
}

@techreport{gelman2013,
 title = "Why ask Why? {F}orward Causal Inference and Reverse Causal Questions",
 author = "Gelman, Andrew and Imbens, Guido",
 institution = "National Bureau of Economic Research",
 type = "Working Paper",
 series = "Working Paper Series",
 number = "19614",
 year = "2013",
 month = "November",
 doi = {10.3386/w19614},
 URL = "http://www.nber.org/papers/w19614"
}

@article{gelman2025,
    author = {Gelman, Andrew and Mikhaeil, Jonas M},
    title = {Russian roulette: the need for stochastic potential outcomes when utilities depend on counterfactuals},
    journal = {Biometrika},
    volume = {112},
    number = {4},
    pages = {asaf062},
    year = {2025},
    month = {08},
    issn = {1464-3510},
    doi = {10.1093/biomet/asaf062},
    url = {https://doi.org/10.1093/biomet/asaf062},
    eprint = {https://academic.oup.com/biomet/article-pdf/112/4/asaf062/63970574/asaf062.pdf},
}

@article{gneezy2000,
  title={A fine is a price},
  author={Gneezy, Uri and Rustichini, Aldo},
  journal={The journal of legal studies},
  volume={29},
  number={1},
  pages={1--17},
  year={2000},
  publisher={The University of Chicago Press}
}

@article{greenland1986,
  title={Identifiability, exchangeability, and epidemiological confounding},
  author={Greenland, Sander and Robins, James M},
  journal={International journal of epidemiology},
  volume={15},
  number={3},
  pages={413--419},
  year={1986},
  publisher={Oxford University Press}
}

@techreport{guggenberger2024,
  title={Minimax regret treatment rules with finite samples when a quantile is the object of interest},
  author={Guggenberger, Patrik and Mehta, Nihal and Pavlov, Nikita},
  year={2024},
  institution={The Pennsylvania State University}
}

@article{heckman1997,
 ISSN = {00346527, 1467937X},
 URL = {http://www.jstor.org/stable/2971729},
 author = {James J. Heckman and Jeffrey Smith and Nancy Clements},
 journal = {The Review of Economic Studies},
 number = {4},
 pages = {487--535},
 publisher = {[Oxford University Press, Review of Economic Studies, Ltd.]},
 title = {Making the Most Out of Programme Evaluations and Social Experiments: Accounting for Heterogeneity in Programme Impacts},
 volume = {64},
 year = {1997}
}

@article{heckman1999,
	title={Local instrumental variables and latent variable models for identifying and bounding treatment effects},
	author={Heckman, James J and Vytlacil, Edward J},
	journal={Proceedings of the National Academy of Sciences},
	volume={96},
	number={8},
	pages={4730--4734},
	year={1999},
	publisher={National Acad Sciences}
}

@article{hirano2008,
  title={Decision theory in econometrics},
  author={Hirano, Keisuke},
  journal={The New Palgrave Dictionary of Economics, 2nd Edition. Eds. S. Durlauf and Le Blume. Palgrave Macmillan},
  year={2008}
}

@incollection{hirano2020,
  title={Asymptotic analysis of statistical decision rules in econometrics},
  author={Hirano, Keisuke and Porter, Jack R},
  booktitle={Handbook of econometrics},
  volume={7},
  pages={283--354},
  year={2020},
  publisher={Elsevier}
}

@article{hirano2009,
  title={Asymptotics for statistical treatment rules},
  author={Hirano, Keisuke and Porter, Jack R},
  journal={Econometrica},
  volume={77},
  number={5},
  pages={1683--1701},
  year={2009},
  publisher={Wiley Online Library}
}

@article{hoeffding1940,
author={Hoeffding, Wassily},
title={Scale-Invariant Correlation Theory},
year={1940},
journal={Schriften des Mathematischen Instituts und des Instituts f{\"u}r Angewandte Mathematik der Universit{\"a}t Berlin},
volume=5,
number={3},
pages={181-233},
note={Translated by Dana Quade in \emph{The Collected Works of Wassily Hoeffding}, ed.  Fisher, N. I.
and Sen, P. K.,  pp. 57--107, New York, NY: Springer New York, 1994.}
}

@article{holland1986,
  title={Statistics and causal inference},
  author={Holland, Paul W},
  journal={Journal of the American Statistical Association},
  volume={81},
  number={396},
  pages={945--960},
  year={1986},
  publisher={Taylor \& Francis}
}

@article{horowitz2000,
  title={Nonparametric analysis of randomized experiments with missing covariate and outcome data},
  author={Horowitz, Joel L and Manski, Charles F},
  journal={Journal of the American statistical Association},
  volume={95},
  number={449},
  pages={77--84},
  year={2000},
  publisher={Taylor \& Francis}
}

@TechReport{huber2012,
  author={Huber, Martin and Mellace, Giovanni},
  title={{Relaxing monotonicity in the identification of local average treatment effects}},
  year=2012,
  month=May,
  institution={University of St. Gallen, School of Economics and Political Science},
  type={Economics Working Paper Series},
  url={https://ideas.repec.org/p/usg/econwp/201212.html},
  number={1212},
  doi={},
}

@article{huber2015testing,
  title={Testing instrument validity for LATE identification based on inequality moment constraints},
  author={Huber, Martin and Mellace, Giovanni},
  journal={Review of Economics and Statistics},
  volume={97},
  number={2},
  pages={398--411},
  year={2015},
  publisher={MIT Press}
}

@article{imbens1994,
 ISSN = {00129682, 14680262},
 URL = {http://www.jstor.org/stable/2951620},
 author = {Guido W. Imbens and Joshua D. Angrist},
 journal = {Econometrica},
 number = {2},
 pages = {467--475},
 publisher = {[Wiley, Econometric Society]},
 title = {Identification and Estimation of Local Average Treatment Effects},
 volume = {62},
 year = {1994}
}

@article{imbens1997,
  title={Estimating outcome distributions for compliers in instrumental variables models},
  author={Imbens, Guido W and Rubin, Donald B},
  journal={The Review of Economic Studies},
  volume={64},
  number={4},
  pages={555--574},
  year={1997},
  publisher={Oxford University Press}
}

@article{imbens2004,
  title={Confidence intervals for partially identified parameters},
  author={Imbens, Guido W and Manski, Charles F},
  journal={Econometrica},
  volume={72},
  number={6},
  pages={1845--1857},
  year={2004},
  publisher={Wiley Online Library}
}

@article{imbens2020,
  title={Potential outcome and directed acyclic graph approaches to causality: Relevance for empirical practice in economics},
  author={Imbens, Guido W},
  journal={Journal of Economic Literature},
  volume={58},
  number={4},
  pages={1129--1179},
  year={2020},
  publisher={American Economic Association 2014 Broadway, Suite 305, Nashville, TN 37203-2425}
}

@article{jaynes1957a,
  title={Information theory and statistical mechanics},
  author={Jaynes, Edwin T},
  journal={Physical review},
  volume={106},
  number={4},
  pages={620},
  year={1957},
  publisher={APS}
}

@article{jaynes1957b,
  title={Information theory and statistical mechanics. II},
  author={Jaynes, Edwin T},
  journal={Physical review},
  volume={108},
  number={2},
  pages={171},
  year={1957},
  publisher={APS}
}

@article{jaynes1968,
  title={Prior probabilities},
  author={Jaynes, Edwin T},
  journal={IEEE Transactions on systems science and cybernetics},
  volume={4},
  number={3},
  pages={227--241},
  year={1968},
  publisher={IEEE}
}

@misc{johnson2003,
  title={Do defaults save lives?},
  author={Johnson, Eric J and Goldstein, Daniel},
  journal={Science},
  volume={302},
  number={5649},
  pages={1338--1339},
  year={2003},
  publisher={American Association for the Advancement of Science}
}

@article{jun2023,
    title={Identifying the Effect of Persuasion},
    author={Jun, Sung Jae and Lee, Sokbae},
    journal={Journal of Political Economy},
    volume={131},
    number={8},
    pages={2032-2058},
    year={2023},
    doi={10.1086/724114},
    URL = {https://doi.org/10.1086/724114}
}

@article{jun2024,
title = {An information–Theoretic approach to partially identified auction models},
journal = {Journal of Econometrics},
volume = {238},
number = {2},
pages = {105566},
year = {2024},
issn = {0304-4076},
doi = {https://doi.org/10.1016/j.jeconom.2023.105566},
url = {https://www.sciencedirect.com/science/article/pii/S0304407623002828},
author = {Jun, Sung Jae and Joris Pinkse}
}

@article{katz2001,
	title={Moving to Opportunity in Boston: Early Results of a Randomized Mobility Experiment},
	author={Katz, Lawrence F and Kling, Jeffrey R and Liebman, Jeffrey B and others},
	journal={The Quarterly Journal of Economics},
	volume={116},
	number={2},
	pages={607--654},
	year={2001},
	publisher={Oxford University Press}
}

@book{kempthorne1952,
  title={Design and Analysis of Experiments},
  author={Kempthorne, O.},
  isbn={9780471468608},
  lccn={51013460},
  year={1952},
  publisher={Wiley},
  address={New York}
}

@article{kessler2025,
  title={Increasing organ donor registration as a means to increase transplantation: an experiment with actual organ donor registrations},
  author={Kessler, Judd B and Roth, Alvin E},
  journal={American Economic Journal: Economic Policy},
  volume={17},
  number={2},
  pages={60--83},
  year={2025},
  publisher={American Economic Association 2014 Broadway, Suite 305, Nashville, TN 37203-2425}
}

@article{kitagawa2015,
  title={A test for instrument validity},
  author={Kitagawa, Toru},
  journal={Econometrica},
  volume={83},
  number={5},
  pages={2043--2063},
  year={2015},
  publisher={Wiley Online Library}
}

@article{kitagawa2018,
  title={Who should be treated? empirical welfare maximization methods for treatment choice},
  author={Kitagawa, Toru and Tetenov, Aleksey},
  journal={Econometrica},
  volume={86},
  number={2},
  pages={591--616},
  year={2018},
  publisher={Wiley Online Library}
}

@article{kline2021,
author = {Kline, Patrick and Walters, Christopher},
title = {Reasonable Doubt: Experimental Detection of Job-Level Employment Discrimination},
journal = {Econometrica},
volume = {89},
number = {2},
pages = {765-792},
doi = {https://doi.org/10.3982/ECTA17489},
url = {https://onlinelibrary.wiley.com/doi/abs/10.3982/ECTA17489},
eprint = {https://onlinelibrary.wiley.com/doi/pdf/10.3982/ECTA17489},
year = {2021}
}

@misc{koch2025,
      title={Statistical Decision Theory with Counterfactual Loss}, 
      author={Benedikt Koch and Kosuke Imai},
      year={2025},
      eprint={2505.08908},
      archivePrefix={arXiv},
      primaryClass={math.ST},
      url={https://arxiv.org/abs/2505.08908}, 
}

@techreport{kowalski2019a,
 title = "Counting Defiers",
 author = "Kowalski, Amanda E",
 institution = "National Bureau of Economic Research",
 type = "Working Paper",
 series = "Working Paper Series",
 number = "25671",
 year = "2019",
 month = "March",
 doi = {10.3386/w25671},
 URL = "http://www.nber.org/papers/w25671"
}

@techreport{kowalski2019b,
 title = "A Model of a Randomized Experiment with an Application to the PROWESS Clinical Trial",
 author = "Kowalski, Amanda E",
 institution = "National Bureau of Economic Research",
 type = "Working Paper",
 series = "Working Paper Series",
 number = "25670",
 year = "2019",
 month = "March",
 doi = {10.3386/w25670},
 URL = "http://www.nber.org/papers/w25670"
}

@article{kowalski2023behavior,
    author = {Kowalski, Amanda E},
    title = {Behaviour within a Clinical Trial and Implications for Mammography Guidelines},
    journal = {The Review of Economic Studies},
    volume = {90},
    number = {1},
    pages = {432-462},
    year = {2023},
    issn = {0034-6527},
    doi = {10.1093/restud/rdac022},
    url = {https://doi.org/10.1093/restud/rdac022},
    eprint = {https://academic.oup.com/restud/article-pdf/90/1/432/48523214/rdac022.pdf},
}

@article{kowalski2023reconciling,
    author = {Kowalski, Amanda E.},
    title = {Reconciling Seemingly Contradictory Results from the {O}regon Health Insurance Experiment and the {M}assachusetts Health Reform},
    journal = {The Review of Economics and Statistics},
    volume = {105},
    number = {3},
    pages = {646-664},
    year = {2023},
    issn = {0034-6535},
    doi = {10.1162/rest_a_01069},
    url = {https://doi.org/10.1162/rest\_a\_01069},
    eprint = {https://direct.mit.edu/rest/article-pdf/105/3/646/2090027/rest\_a\_01069.pdf},
}

@incollection{kuhn1953,
	url = {https://doi.org/10.1515/9781400881970-012},
	title = {Extensive Games and the Problem of Information},
	booktitle = {Contributions to the Theory of Games, Volume II},
	author = {H. W. Kuhn},
	editor = {Harold William Kuhn and Albert William Tucker},
	publisher = {Princeton University Press},
	address = {Princeton},
	pages = {193--216},
	doi = {doi:10.1515/9781400881970-012},
	isbn = {9781400881970},
	year = {1953},
	lastchecked = {2024-12-14}
}

@article{lacetera2010,
  title={Do all material incentives for pro-social activities backfire? The response to cash and non-cash incentives for blood donations},
  author={Lacetera, Nicola and Macis, Mario},
  journal={Journal of Economic Psychology},
  volume={31},
  number={4},
  pages={738--748},
  year={2010},
  publisher={Elsevier}
}

@article{li2016,
  title={Exact confidence intervals for the average causal effect on a binary outcome},
  author={Li, Xinran and Ding, Peng},
  journal={Statistics in Medicine},
  volume={35},
  number={6},
  pages={957--960},
  year={2016},
  publisher={Wiley Online Library}
}

@inproceedings{li2019,
  title={Unit selection based on counterfactual logic},
  author={Li, Ang and Pearl, Judea},
  booktitle={Proceedings of the Twenty-Eighth International Joint Conference on Artificial Intelligence},
  year={2019}
}

@book{lindley1985making,
  title={Making Decisions},
  author={Lindley, D.V.},
  isbn={9780471908036},
  lccn={lc85012010},
  url={https://books.google.com/books?id=3-ZQAAAAMAAJ},
  year={1985},
  publisher={Wiley}
}

@Book{list2026,
publisher={University of Chicago Press},
series={University of Chicago Press Economics Books},
edition={1},
author={List, John A.},
title={Experimental Economics},
year={2026},
month={September},
number={9780226820651},
volume={None},
doi={None},
url={https://ideas.repec.org/b/ucp/bkecon/9780226820651.html},
}

@article{machado2019,
  title={Instrumental variables and the sign of the average treatment effect},
  author={Machado, Cecilia and Shaikh, Azeem M and Vytlacil, Edward J},
  journal={Journal of Econometrics},
  year={2019},
  volume={212},
  issue={2},
  pages={522--555},
  publisher={Elsevier}
}

@article{manski1992,
  author = {Charles F. Manski and Gary D. Sandefur and Sara McLanahan and Daniel Powers},
  title = {Alternative Estimates of the Effect of Family Structure during Adolescence on High School Graduation},
  journal = {Journal of the American Statistical Association},
  volume = {87},
  number = {417},
  pages = {25--37},
  year = {1992},
  publisher = {ASA Website},
  doi = {10.1080/01621459.1992.10475171},
  URL = {https://www.tandfonline.com/doi/abs/10.1080/01621459.1992.10475171},
  eprint = {https://www.tandfonline.com/doi/pdf/10.1080/01621459.1992.10475171}
}

@article{manski1997mixing,
  title={The mixing problem in programme evaluation},
  author={Manski, Charles F},
  journal={The Review of Economic Studies},
  volume={64},
  number={4},
  pages={537--553},
  year={1997},
  publisher={Wiley-Blackwell}
}

@article{manski1997monotone,
  title={Monotone treatment response},
  author={Manski, Charles F},
  journal={Econometrica},
  volume={65},
  number={6},
  pages={1311--1334},
  year={1997},
  publisher={JSTOR}
}

@article{manski2004,
  title={Statistical treatment rules for heterogeneous populations},
  author={Manski, Charles F},
  journal={Econometrica},
  volume={72},
  number={4},
  pages={1221--1246},
  year={2004},
  publisher={Wiley Online Library}
}

@article{manski2007,
  title={Minimax-regret treatment choice with missing outcome data},
  author={Manski, Charles F},
  journal={Journal of Econometrics},
  volume={139},
  number={1},
  pages={105--115},
  year={2007},
  publisher={Elsevier}
}

@article{manski2007admissible,
  title={Admissible treatment rules for a risk-averse planner with experimental data on an innovation},
  author={Manski, Charles F and Tetenov, Aleksey},
  journal={Journal of Statistical Planning and Inference},
  volume={137},
  number={6},
  pages={1998--2010},
  year={2007},
  publisher={Elsevier}
}

@article{manski2018,
  title={Reasonable patient care under uncertainty},
  author={Manski, Charles F},
  journal={Health Economics},
  volume={27},
  number={10},
  pages={1397--1421},
  year={2018},
  publisher={Wiley Online Library}
}

@article{manski2019,
author = {Charles F. Manski},
title = {Treatment Choice With Trial Data: Statistical Decision Theory Should Supplant Hypothesis Testing},
journal = {The American Statistician},
volume = {73},
number = {sup1},
pages = {296-304},
year  = {2019},
publisher = {Taylor & Francis},
doi = {10.1080/00031305.2018.1513377},
URL = {https://doi.org/10.1080/00031305.2018.1513377},
eprint = {https://doi.org/10.1080/00031305.2018.1513377}
}

@article{manski2021,
  title={Statistical Decision Properties of Imprecise Trials Assessing Coronavirus Disease 2019 (COVID-19) Drugs},
  author={Manski, Charles F and Tetenov, Aleksey},
  journal={Value in Health},
  volume={24},
  number={5},
  pages={641--647},
  year={2021},
  publisher={Elsevier}
}

@article{mellstrom2008,
  title={Crowding out in blood donation: was Titmuss right?},
  author={Mellstr{\"o}m, Carl and Johannesson, Magnus},
  journal={Journal of the European Economic Association},
  volume={6},
  number={4},
  pages={845--863},
  year={2008},
  publisher={Oxford University Press}
}

@Article{mueller2023,
journal={Journal of Causal Inference},
author={Mueller, Scott and Pearl, Judea},
title={Personalized decision making – A conceptual introduction},
year={2023},
month={January},
pages={1-13},
volume={11},
number={1},
doi={10.1515/jci-2022-0050},
url={https://ideas.repec.org/a/bpj/causin/v11y2023i1p13n1.html},
}

@article{mourifie2017,
  title={Testing local average treatment effect assumptions},
  author={Mourifi{\'e}, Ismael and Wan, Yuanyuan},
  journal={Review of Economics and Statistics},
  volume={99},
  number={2},
  pages={305--313},
  year={2017},
  publisher={MIT Press}
}

@article{mullahy2018,
title = "Individual results may vary: Inequality-probability bounds for some health-outcome treatment effects",
journal = "Journal of Health Economics",
volume = "61",
pages = "151 - 162",
year = "2018",
issn = "0167-6296",
doi = "https://doi.org/10.1016/j.jhealeco.2018.06.011",
url = "http://www.sciencedirect.com/science/article/pii/S0167629617309463",
author = "John Mullahy"
}

@article{neyman1923,
  title={On the Application of Probability Theory to Agricultural Experiments. {E}ssay on Principles. {S}ection 9},
  author={Neyman, Jersey},
  journal={Roczniki Nauk Rolniczych},
  volume={10},
  pages={1--51},
  year={1923},
  note={Translated by D.M. Dabrowski and T.P. Speed in \emph{Statistical Science 5}(4), pp. 465--472, 1990.}
}

@article{pearl1999,
 ISSN = {00397857, 15730964},
 URL = {http://www.jstor.org/stable/20118223},
 author = {Judea Pearl},
 journal = {Synthese},
 number = {1/2},
 pages = {93--149},
 publisher = {Springer},
 title = {Probabilities of Causation: Three Counterfactual Interpretations and Their Identification},
 volume = {121},
 year = {1999}
}

@book{pearl2018,
  title={The Book of Why: The New Science of Cause and Effect},
  author={Pearl, Judea and Mackenzie, Dana},
  year={2018},
  publisher={Basic books}
}

@article{permutt1989,
  title={Simultaneous-equation estimation in a clinical trial of the effect of smoking on birth weight},
  author={Permutt, Thomas and Hebel, J Richard},
  journal={Biometrics},
  pages={619--622},
  year={1989},
  publisher={JSTOR}
}

@article{richardson2010,
  title={Analysis of the binary instrumental variable model},
  author={Richardson, Thomas S and Robins, James M},
  journal={Heuristics, Probability and Causality: A Tribute to Judea Pearl},
  pages={415--444},
  year={2010},
  publisher={College Publications London, United Kingdom}
}

@article{rigdon2015,
  title={Randomization inference for treatment effects on a binary outcome},
  author={Rigdon, Joseph and Hudgens, Michael G},
  journal={Statistics in Medicine},
  volume={34},
  number={6},
  pages={924--935},
  year={2015},
  publisher={Wiley Online Library}
}

@article{rubin1974,
  title={Estimating causal effects of treatments in randomized and nonrandomized studies.},
  author={Rubin, Donald B},
  journal={Journal of Educational Psychology},
  volume={66},
  number={5},
  pages={688--701},
  year={1974},
  publisher={American Psychological Association}
}

@article{rubin1977,
  title={Assignment to Treatment Group on the Basis of a Covariate},
  author={Rubin, Donald B},
  journal={Journal of Educational and Behavioral Statistics},
  volume={2},
  number={1},
  pages={1--26},
  year={1977},
  publisher={Sage Publications}
}

@article{rubin1980,
 ISSN = {01621459},
 URL = {http://www.jstor.org/stable/2287653},
 author = {Donald B. Rubin},
 journal = {Journal of the American Statistical Association},
 number = {371},
 pages = {591--593},
 publisher = {[American Statistical Association, Taylor & Francis, Ltd.]},
 title = {Randomization Analysis of Experimental Data: The {Fisher Randomization Test c}omment},
 volume = {75},
 year = {1980}
}

@article{schlag2003,
  title={How to Minimize Maximum Regret in Repeated Decision Making},
  author={Schlag, Karl H},
  journal={Unpublished Manuscript, European University Institute},
  year={2003}
}

@article{schlag2007,
  title={Eleven - Designing Randomized Experiments under Minimax Regret},
  author={Schlag, Karl H},
  journal={Unpublished manuscript, European University Institute, Florence},
  year={2007}
}

@article{schneider2023,
  title={Financial incentives for vaccination do not have negative unintended consequences},
  author={Schneider, Florian H and Campos-Mercade, Pol and Meier, Stephan and Pope, Devin and Wengstr{\"o}m, Erik and Meier, Armando N},
  journal={Nature},
  volume={613},
  number={7944},
  pages={526--533},
  year={2023},
  publisher={Nature Publishing Group UK London}
}

@article{semenova2024,
      title={Aggregated Intersection Bounds and Aggregated Minimax Values}, 
      author={Vira Semenova},
      year={2024},
      journal={arXiv preprint arXiv:2303.00982} 
}

@article{stoye2007,
  title={Minimax regret treatment choice with incomplete data and many treatments},
  author={Stoye, J{\"o}rg},
  journal={Econometric Theory},
  volume={23},
  number={1},
  pages={190--199},
  year={2007},
  publisher={Cambridge University Press}
}

@article{stoye2009,
  title={Minimax regret treatment choice with finite samples},
  author={Stoye, J{\"o}rg},
  journal={Journal of Econometrics},
  volume={151},
  number={1},
  pages={70--81},
  year={2009},
  publisher={Elsevier}
}

@article{stoye2012,
  title={Minimax regret treatment choice with covariates or with limited validity of experiments},
  author={Stoye, J{\"o}rg},
  journal={Journal of Econometrics},
  volume={166},
  number={1},
  pages={138--156},
  year={2012},
  publisher={Elsevier}
}

@article{tappin2015,
  title={Financial incentives for smoking cessation in pregnancy: randomised controlled trial},
  author={Tappin, David and Bauld, Linda and Purves, David and Boyd, Kathleen and Sinclair, Lesley and MacAskill, Susan and McKell, Jennifer and Friel, Brenda and McConnachie, Alex and De Caestecker, Linda and others},
  journal={Bmj},
  volume={350},
  year={2015},
  publisher={British Medical Journal Publishing Group}
}

@inproceedings{tamer2004,
  title={Parameter Set Inference in a Class of Econometric Models},
  author={Tamer, Elie and Chernozhukov, Victor and Hong, Han},
  booktitle={Econometric Society 2004 North American Winter Meetings},
  number={382},
  year={2004},
  organization={Econometric Society}
}

@article{tchetgen2024,
  title={The Nudge Average Treatment Effect},
  author={Tchetgen Tchetgen, Eric J},
  journal={arXiv preprint arxiv:2410.23590},
  year={2024}
}

@article{tetenov2012,
  title={Statistical treatment choice based on asymmetric minimax regret criteria},
  author={Tetenov, Aleksey},
  journal={Journal of Econometrics},
  volume={166},
  number={1},
  pages={157--165},
  year={2012},
  publisher={Elsevier}
}

@article{tian2000,
  title={Probabilities of causation: Bounds and identification},
  author={Tian, Jin and Pearl, Judea},
  journal={Annals of Mathematics and Artificial Intelligence},
  volume={28},
  number={1-4},
  pages={287--313},
  year={2000},
  publisher={Springer}
}

@article{vytlacil2002,
  title={Independence, monotonicity, and latent index models: An equivalence result},
  author={Vytlacil, Edward},
  journal={Econometrica},
  volume={70},
  number={1},
  pages={331--341},
  year={2002},
  publisher={Wiley Online Library}
}

@article{wager2018,
	author = {Stefan Wager and Susan Athey},
	title = {Estimation and Inference of Heterogeneous Treatment Effects using Random Forests},
	journal = {Journal of the American Statistical Association},
	volume = {113},
	number = {523},
	pages = {1228-1242},
	year  = {2018},
	publisher = {Taylor & Francis},
	doi = {10.1080/01621459.2017.1319839},
	URL = {https://doi.org/10.1080/01621459.2017.1319839},
	eprint = {https://doi.org/10.1080/01621459.2017.1319839}
}

@article{wald1949,
	author = {Abraham Wald},
	title = {{Statistical Decision Functions}},
	volume = {20},
	journal = {The Annals of Mathematical Statistics},
	number = {2},
	publisher = {Institute of Mathematical Statistics},
	pages = {165 -- 205},
	year = {1949},
	doi = {10.1214/aoms/1177730030},
	URL = {https://doi.org/10.1214/aoms/1177730030}
}

@article{wang2016,
  title={Maximum Entropy and Bayesian Inference for the Monty Hall Problem},
  author={Wang, Jennifer L and Tran, Tina and Abebe, Fisseha},
  journal={Journal of Applied Mathematics and Physics},
  volume={4},
  number={7},
  pages={1222--1230},
  year={2016},
  publisher={Scientific Research Publishing}
}

@article{welch1937,
  title={On the z-test in randomized blocks and Latin squares},
  author={Welch, Bernard L},
  journal={Biometrika},
  volume={29},
  number={1/2},
  pages={21--52},
  year={1937},
  publisher={JSTOR}
}

@article{young2019,
  title={{Channeling Fisher:} Randomization tests and the statistical insignificance of seemingly significant experimental results},
  author={Young, Alwyn},
  journal={The Quarterly Journal of Economics},
  volume={134},
  number={2},
  pages={557--598},
  year={2019},
  publisher={Oxford University Press}
}

@article{zhang2003,
author = {Junni L. Zhang and Donald B. Rubin},
title ={Estimation of Causal Effects via Principal Stratification When Some Outcomes are Truncated by ``Death"},
journal = {Journal of Educational and Behavioral Statistics},
volume = {28},
number = {4},
pages = {353-368},
year = {2003},
doi = {10.3102/10769986028004353},
URL = {https://doi.org/10.3102/10769986028004353},
eprint = {https://doi.org/10.3102/10769986028004353},
}

\end{document}